\documentclass[preprint,12pt]{elsarticle}
\biboptions{sort&compress}

\usepackage{verbatim} 
\usepackage{epsfig}

\usepackage{amssymb} 

\usepackage{mathtools}

\usepackage{lineno}

\usepackage{tikz}        
\usepackage{etoolbox}

\usepackage{soul}  

\usepackage[caption=false]{subfig}

\begin{document}

\begin{frontmatter}



\title{Microscopic dynamics of the evacuation
phenomena}

\author[label1]{F.E.~Cornes}
\author[label3]{G.A.~Frank}
\author[label1,label2]{C.O.~Dorso}
\address[label1]{Departamento de F\'\i sica, Facultad de Ciencias 
Exactas y Naturales, Universidad de Buenos Aires, Pabell\'on I, Ciudad 
Universitaria, 1428 Buenos Aires, Argentina.}
\address[label2]{Instituto de F\'\i sica de Buenos Aires, Pabell\'on I, Ciudad 
Universitaria, 1428 Buenos Aires, Argentina.}
\address[label3]{ Unidad de Investigaci\'on y Desarrollo de las 
Ingenier\'\i as, Universidad Tecnol\'ogica Nacional, Facultad Regional Buenos 
Aires, Av. Medrano 951, 1179 Buenos Aires, Argentina.}



\begin{abstract}


We studied the room evacuation problem within the context of the Social Force 
Model. We focused on a system of 225 pedestrians escaping from a room in 
different anxiety levels, and analyzed the clogging delays as the relevant 
magnitude responsible for the evacuation performance. We linked the delays with 
the clusterization phenomenon along the \textit{faster is slower} and the 
\textit{faster is faster} regimes. We will show that the \textit{faster is 
faster} regime is characterized by the presence of a giant cluster structure 
(composed by more than 15 pedestrians), although no long lasting delays appear 
within this regime. For this system, we found that the relevant structures in 
the \textit{faster is slower} regime are those blocking clusters that are 
somehow attached to the two walls defining the exit. At very low desired 
velocities, very small structures become relevant (composed by less than 5 
pedestrians), but at intermediate velocities ($v_d\simeq3\,$m/s) the pedestrians 
involved in the blockings increases (not exceeding 15 pedestrians). 


\end{abstract}


\end{frontmatter}


\section{\label{introduction}Introduction}

The problem of emergency evacuations has become relevant in the last
decades due to the increasing number and size of mass events
(religious events, music festivals, etc). The evacuation through
narrow pathways or doorways appears as one of the most problematic scenarios in 
the literature. Many experiments and numerical simulations
have been carried out for a better understanding of the crowd behavior in a
variety of situations. The door width \cite{Haghani_2019,Sarvi_2019, 
Boltes-Seyfried_2018,Zuriguel_2016,Seyfried-Boltes_2014,Hoogendoorn_2012,
Seyfried-Boltes_2011,Boltes_2009,Seyfried-Schadschneider_2009}, the crowd 
composition \cite{Hoogendoorn_2012, Kuperman_2017}, or even the position of the 
exit \cite{Sarvi_2019, Shiwatoki_2016} raise as relevant magnitudes affecting 
the evacuation performance. The environmental and contextual realism, the 
pressure level, the occupancy density, the sample size, etc. are some inherent 
limitations of their experiments. Thus, the comparison between the results of 
the controlled experiments and the emergency escape behavior is quite 
controversial \cite{Haghani_2019}. \\

Ethical requirements in experimental setting impedes reproduce a
real emergency situation. In this sense, the form that experimental researchers 
simulates a competitive scenario differs from one to another experiment. 
Threfore, many controlled experiments on the pedestrians' egress 
time show dissimilar results. For instance, Refs.~\cite{Zuriguel_2016, 
Zuriguel_2015, Zuriguel_2014} report that under a simulated competitive 
scenario (say, allowing the contact between the pedestrians), the more the 
pedestrian's anxiety to reach the exit, the greater the evacuation time. But 
other experimental reports provide evidence of decreasing evacuation times for 
increasing escaping desires \cite{Haghani_2019, Hoogendoorn_2012}. The former 
correspond to a ``faster-is-slower'' behavior, while the latter correspond to a 
``faster-is-faster'' behavior. Whether the dominant behavior is 
``faster-is-slower'' or ``faster-is-faster'' seems to be a matter of the 
pushing aggressiveness, in spite of the desire to reach the 
exit \cite{Haghani_2019}. Thus, it has been proposed recently that the term 
``faster-is-slower'' should be replaced by the more precise one: ``aggressive 
egress-is-slower''. \\

The Social Force Model (SFM) introduced by Helbing \cite{helbing_2000} was the 
first model to accomplish a ``faster-is-slower'' effect for moderate to high 
anxiety levels within the crowd. This occurs whenever friction dominates the
pedestrian dynamic \cite{dorso_2007, dorso_2005, dorso_2017}. The 
``faster-is-faster'' phenomenon was also reported within this model, but for 
very high pressures among the individuals \cite{dorso_2017}.\\

The time-lapse between consecutive evacuees and the egress curve are one of 
the most common measures for experimental and computational research 
\cite{Hoogendoorn_2012, Seyfried-Boltes_2011, Boltes_2009, Shiwatoki_2016, 
Zuriguel_2015, Zuriguel_2014, dorso_2005, dorso_2017}. The egress curve shows 
that the leaving flow may occur either regularly or intermittently, depending on 
the pedestrians' anxiety to reach the exit. Clogging can be observed as 
``arching structures'' just before any narrowing of the leaving pathway 
\cite{helbing_2000, dorso_2005, dorso_2017, Sticco_2017}. Clogging has also been 
related to the pressure inside the \textit{bulk} on this system 
\cite{helbing_2000, dorso_2005, dorso_2017, Zuriguel_2013, Daniels_2011, 
Zuriguel_2013_b, Daniels_2016_a, Daniels_2016_b, Daniels_2018}.\\

Our aim is to analyze the relation between the delays of consecutive leaving 
pedestrians and the topological structures among the crowd during the 
\textit{faster is slower} and \textit{faster is faster} regimes. The work is 
organized as follows. In Section~\ref{sec:background} we introduce the dynamics 
equations for evacuating pedestrians, in the context of the Social Force Model 
(SFM). We also define the meaning of spatial 
clusters. Section~\ref{sec:simulations} details the simulation procedures used 
to studying the room evacuation of a crowd under increasing anxiety levels. 
Section~\ref{sec:results} displays the result of our investigation, while the 
conclusions are summarize in Section~\ref{sec:conclusions}.\\

\section{\label{sec:background}Theoretical background}

\subsection{\label{sec:sfm}The Social Force Model (SFM)}

Our research was carried out in the context of the ``social force model'' (SFM) 
proposed by Helbing and co-workers \cite{helbing_2000}. This model states that
human motion is caused by the desire of people to reach a certain destination 
at a certain velocity, as well as other environmental factors. It is a 
generalized force model, including socio-psychological forces, as well as 
``physical'' forces like friction. These forces enter the Newton equation as 
follows.   

\begin{equation}
m_i\,\displaystyle\frac{d\mathbf{v}^{(i)}}{dt}=\mathbf{f}_d^{(i)}
+\displaystyle\sum_{j=1}^{N}\displaystyle\mathbf{f}_s^{(ij)}
+\displaystyle\sum_ {
j=1}^{N}\mathbf{f}_g^{(ij)}\label{eq_mov}
\end{equation}

\noindent where the $i,j$ subscripts correspond to any two pedestrians in the 
crowd. $\mathbf{v}^{(i)}(t)$ means the current velocity of the pedestrian  
$(i)$, while $\mathbf{f}_d$ and $\mathbf{f}_s$ correspond to the ``desired 
force'' and the ``social force'', respectively. $\mathbf{f}_g$ is the granular 
force. \\

The ``desired force'' $\mathbf{f}_d$ describes the pedestrians own desire to 
move at the desired velocity $v_d$. But, due to 
environmental factors (\textit{i.e.} obstacles, visibility), he (she) actually 
moves at the current velocity $\mathbf{v}^{(i)}(t)$. Thus, he (she) will 
accelerate (or decelerate) to reach the desired velocity $v_d$ that will make 
him (her) feel more comfortable. Thus, in the Social Force Model, the desired 
force reads \cite{helbing_2000}. \\

\begin{equation}
        \mathbf{f}_d^ {(i)}(t) =  
m_i\,\displaystyle\frac{v_d^{(i)}\,\mathbf{e}_d^
{(i)}(t)-\mathbf{v}^{(i)}(t)}{\tau} \label{desired}
\end{equation}

\noindent where $m_i$ is the mass of the pedestrian $i$ and $\tau$ represents 
the relaxation time needed to reach his (her) desired velocity. 
$\mathbf{e}_d$ is the unit vector pointing to the target position. For 
simplicity, we assume that $v_d$ remains constant during an evacuation process, 
but $\mathbf{e}_d$ changes according to the current position of the 
pedestrian. Detailed values for $m_i$ and $\tau$ can be found in 
Refs.~\cite{dorso_2011,helbing_2000}. \\

The ``social force'' $\mathbf{f}_s^{(ij)}$ represents the psychological tendency 
of any two pedestrians, say $i$ and $j$, to stay away from each other 
(\textit{private sphere} preservation). It is represented by a repulsive 
interaction force

\begin{equation}
        \mathbf{f}_s^{(ij)} = A_i\,e^{(R_{ij}-r_{ij})/B_i}\mathbf{n}_{ij} 
        \label{social}
\end{equation}

\noindent where $(ij)$ represents any pedestrian-pedestrian pair, or 
pedestrian-wall pair. $A_i$ and $B_i$ are two fixed parameters (see 
Ref.~\cite{dorso_2005}). The distance $R_{ij}=R_i+R_j$ is the sum of the 
pedestrians radius, while $r_{ij}$ is the distance between the center of mass 
of the pedestrians $i$ and $j$. $\mathbf{n}_{ij}$ stands for the unit vector in 
the $\vec{ji}$ direction. For the case of repulsive feelings with the walls, 
$d_{ij}$ corresponds to the shortest distance between the pedestrian and the 
wall \cite{helbing_2000,helbing_1995}. \\

Any two pedestrians touch each other if their mutual distance $r_{ij}$ is 
smaller than $R_{ij}$. Also, any pedestrian touches a wall if his (her) distance
$r_{ij}$ to the wall \textit{j} is smaller than $R_i$. In these cases, an 
additional force is included in the model, called the “granular force” 
$\mathbf{f}_g$. It is composed by two forces: a sliding friction and a body 
force. The expression for this force is 

\begin{equation}
 \mathbf{f}_g^{(ij)}= \mathbf{f}_{sliding} + \mathbf{f}_{body} 
= \kappa_t\,g(R_{ij}-r_{ij})\,
(\Delta\mathbf{v}^{(ij)}\cdot\hat{\mathbf{t}}_{ij})\,\hat{\mathbf{t}}_{ij}+
\kappa_n\,g(R_{ij}-r_{ij})\,
\,\hat{\mathbf{n}}_{ij}\label{eqn_friction}
\end{equation}

\noindent where $\kappa_t$ and $\kappa_n$ are fixed parameters. The function 
$g(R_{ij}-r_{ij})$ is equal to its argument if $R_{ij}>r_{ij}$ (\textit{i.e.} 
if pedestrians are in contact) and zero for any other 
case. $\Delta\mathbf{v}^{(ij)}\cdot\hat{\mathbf{t}}_{ij}$ 
represents the relative tangential velocities of the sliding bodies (or between 
the individual and the walls).    \\

Notice that the sliding friction occurs in the tangential direction while the 
body force occurs in the normal direction. Both are assumed to be linear with 
respect to the net distance between contacting pedestrians. The coefficients 
$k_t$ (for the sliding friction) and $k_n$ (for the 
body force) are supposed to be related to the areas of contact and the clothes 
material, among others. \\

\subsection{\label{sec:definitions}Some definitions}

In this section we define magnitudes which we have found useful in order to 
explore the system properties. We will focus our attention in two quantities

\begin{itemize}
 \item In Section~\ref{sec:clustering_structures} we will define and 
characterize the concept of clustering structures.
 \item In Section~\ref{sec:def_delay} we will introduce the meaning of a 
clogging delay and we will explain different kinds of it.
 \item In Section~\ref{sec:cac} we will relate the clustering structures 
with clogging delays.\\
\end{itemize}

\subsubsection{\label{sec:clustering_structures}Clustering structures}

Human clustering arises when pedestrians get in contact between 
each other. These morphological structures are responsible for the time delays 
during the evacuation process \cite{dorso_2007,dorso_2005,dorso_2017}. Thus, 
for future analysis a precise definition of this kind of structures is 
needed.

\subsubsection*{\label{sec:def_gc} Clusters}

Many previous works showed that clusters of pedestrians play a fundamental 
role during the evacuation process \cite{dorso_2005,dorso_2007,dorso_2017}. In 
this sense, researchers demonstrated that there exist a relation between 
them and the clogging up of people. Clusters of pedestrians can be defined as 
the set of individuals that for any member of the group (say, $i$) there exists 
at least another member belonging to the same group ($j$) in contact with the 
former. Thus, we define a \textit{spatial cluster} ($C_g$) following the 
mathematical formula given in Ref.~\cite{strachan}. 

\begin{equation}
C_g:p_i~\epsilon~ C_g \Leftrightarrow \exists~ p_j~\epsilon~C_g / r_{ij} < 
(R_i+R_j) \label{ec-cluster}
\end{equation}

\noindent where ($p_i$) indicate the \textit{ith} pedestrian and $R_i$ is his 
(her) radius (shoulder width). This means that $C_g$ is a set of pedestrians 
that interact not only with the social force, but also with physical forces 
(\textit{i.e.} friction force and body force).\\

\begin{figure*}[!ht]
\subfloat[\label{fig:granular_clusters_diag}]{
\includegraphics[scale=0.38]{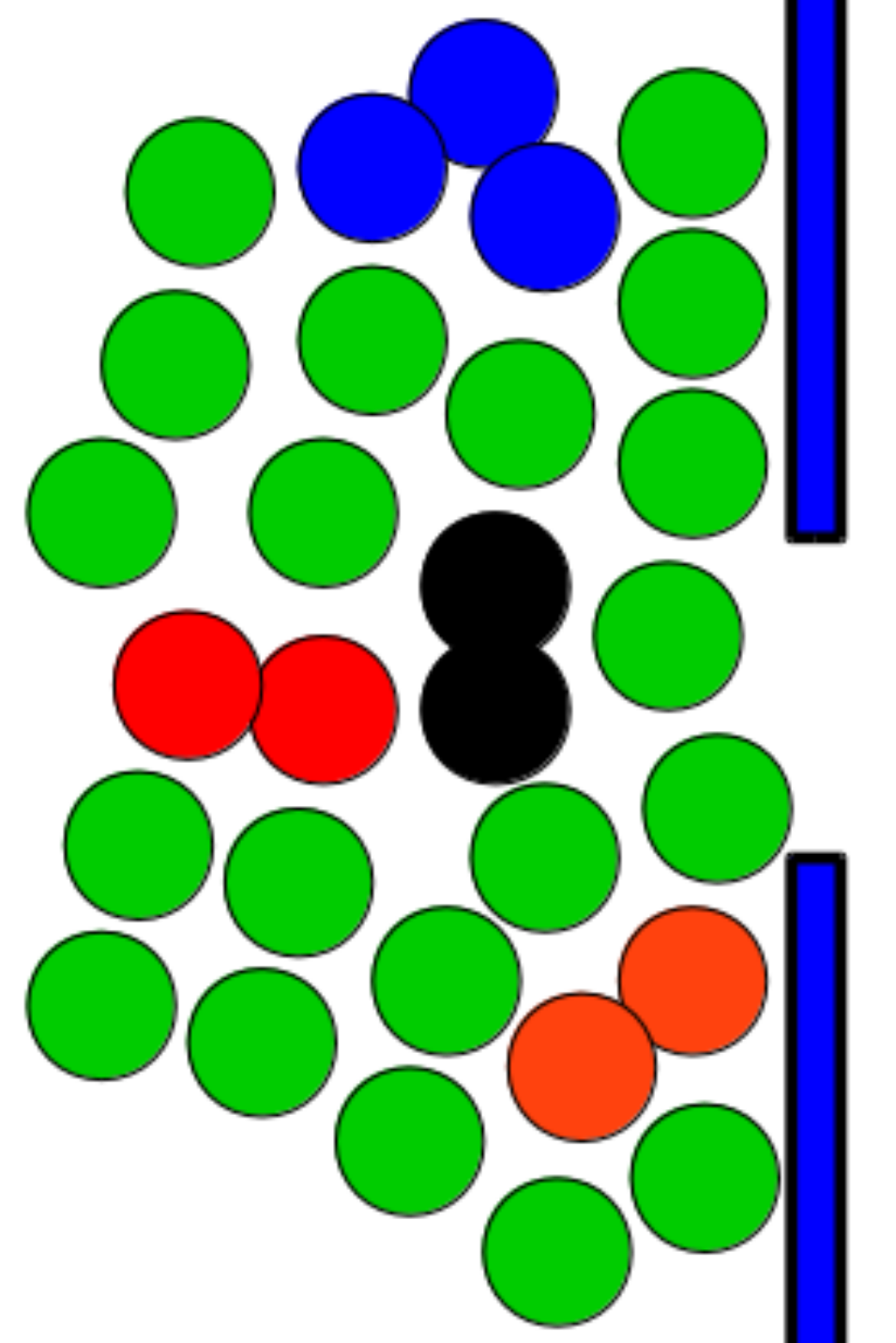}
}
\hspace{3mm}
\subfloat[\label{fig:blocking_cluster_diag}]{
\includegraphics[scale=0.38]{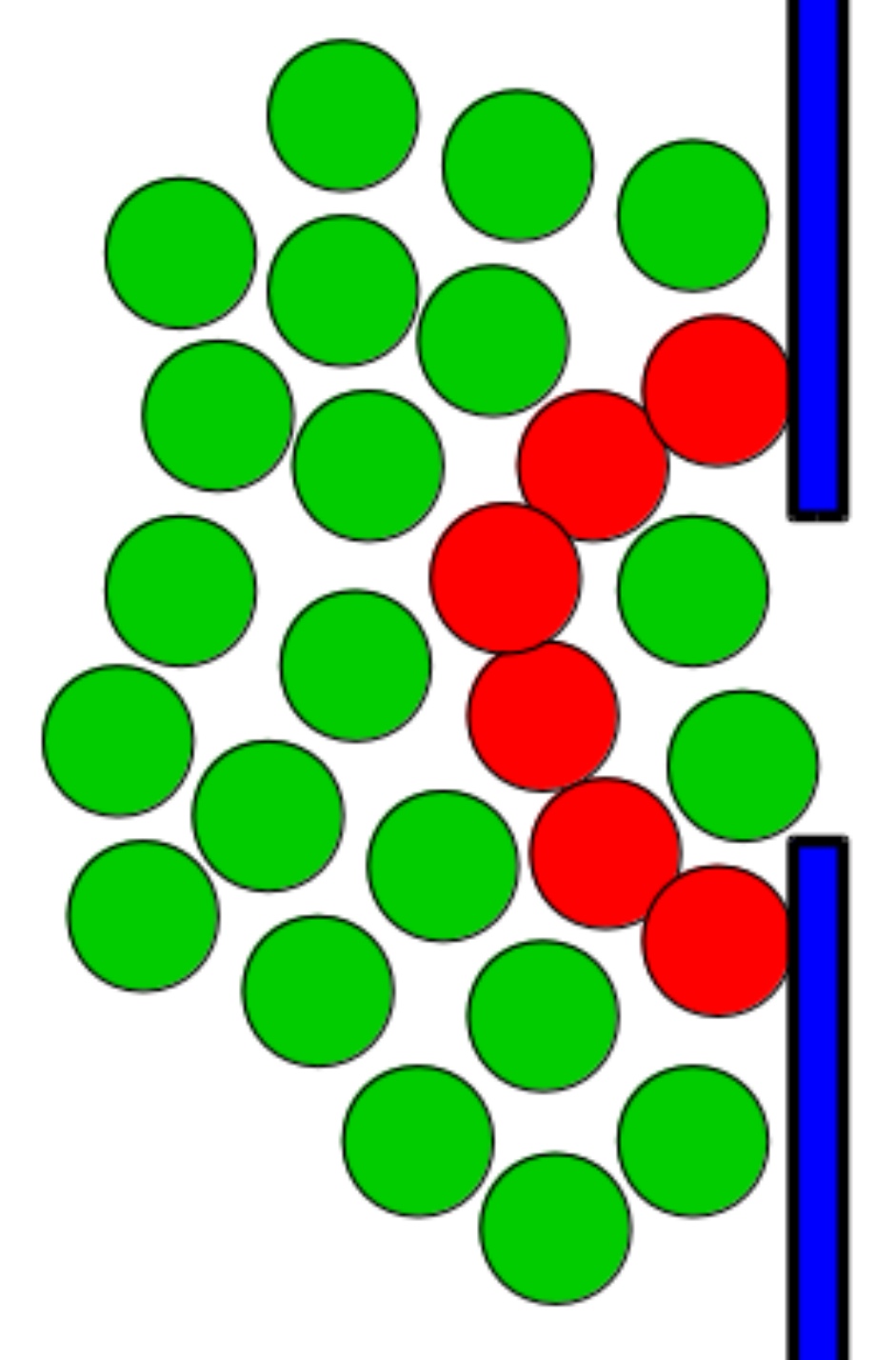}
}
\hspace{3mm}
\subfloat[\label{fig:multiple_diag}]{
\includegraphics[scale=0.38]{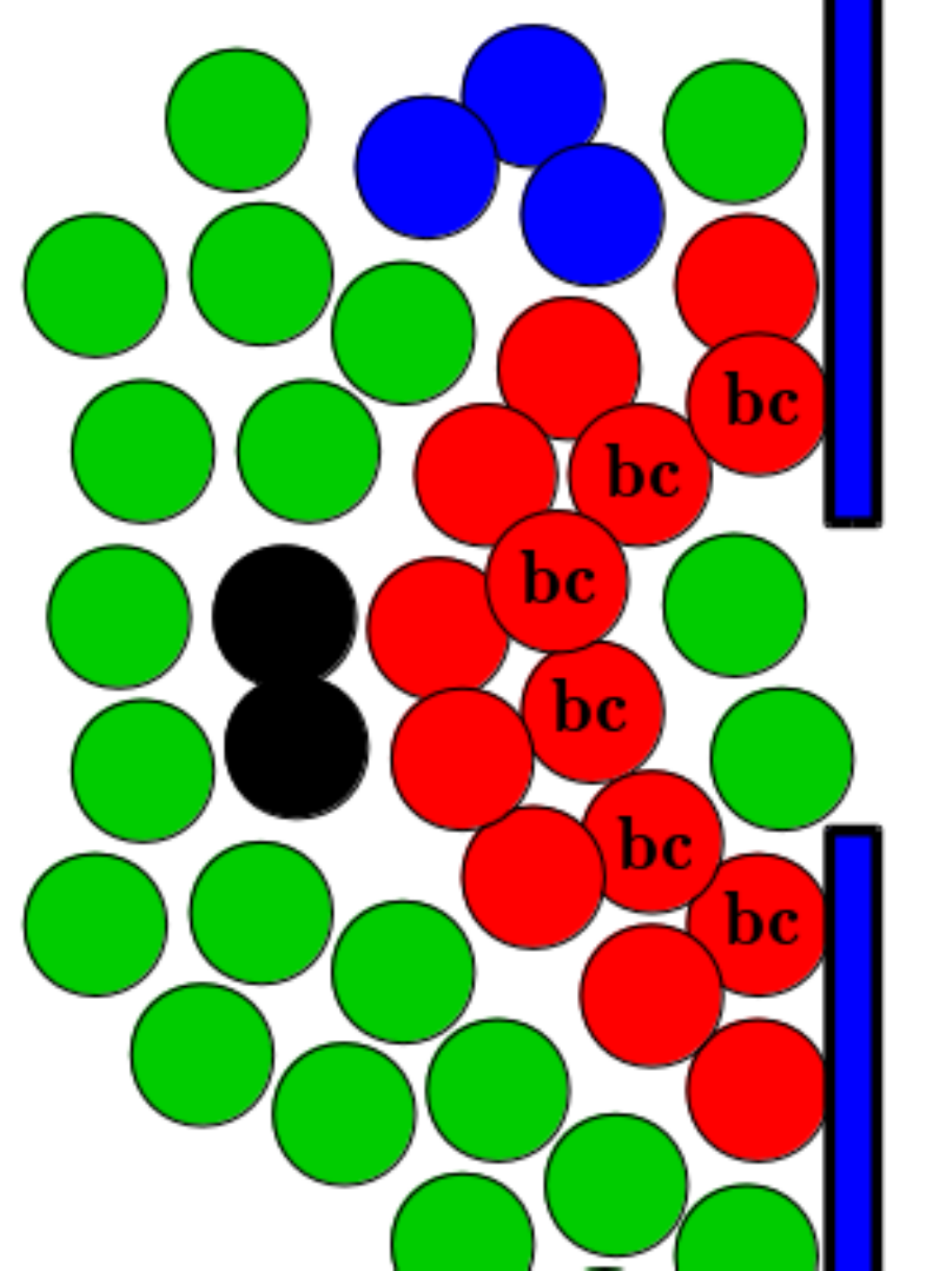}
}
\caption{\label{fig:diagramas_structures} (a,b,c) Schematic representation of 
different spatial structures clusters. Each color represent a different 
spatial cluster. Isolated pedestrians are represented by green circles 
(1-spatial cluster). (a) Three 2-spatial cluster are represented by black, 
orange and red circles. One 3-spatial cluster is represented by blue circles. 
(b) One 6-spatial cluster is represented by red circles. This one corresponds 
to a blocking cluster. (c) Multiple spatial clusters founded through the 
simulations. 3-spatial cluster, 2-spatial cluster and 14-spatial cluster are 
represented by blue, black and red circles, respectively. Also, the 14-spatial 
cluster includes a blocking cluster (labeled as bc). The blue boxes 
represent the walls.} 
\end{figure*}

Fig.~\ref{fig:granular_clusters_diag} shows four spatial clusters. We will label 
a spatial cluster composed by $n$ pedestrians as a $n$-\textit{spatial cluster}. 
Isolated pedestrians (\textit{i.e.} without contact with their surrounding 
individuals) corresponds to a $1$-\textit{spatial cluster} (green circles in 
Fig.~\ref{fig:granular_clusters_diag}).

\subsubsection*{\label{sec:def_bc}Blocking clusters}

The simulations show that some spatial clusters are able to \textit{block} the 
door. We will call \textit{blocking cluster} to these kind of spatial clusters. 
In this sense, a ``blocking cluster'' is defined as the minimum subset of 
clusterized particles (\textit{i.e.} spatial cluster) closest to the door whose 
first and last component particles are in contact with the walls at both sides 
of the door \cite{dorso_2005}. \\

Fig.~\ref{fig:blocking_cluster_diag} shows a typical blocking cluster composed 
by 6 pedestrians (red circles). As a special type of a spatial cluster, these 
can be composed by different number of pedestrians. Our simulations show that 
they can include from 5 to a maximum of 15 pedestrians for a door width of 
0.92~m. Ref.~\cite{dorso_2007} shows that this depends on door 
width.\\

A blocking cluster may coexist simultaneously with other spatial clusters. In 
this sense, we can observe in Fig.~\ref{fig:multiple_diag} three spatial 
clusters: 2, 3 and 14-spatial cluster are present in it. Notice that 
the blocking cluster belongs to a greater spatial cluster of 14 pedestrians. As 
we will see in Section \ref{sec:clusters}, this occurs more frequently for high 
desired velocities. \\

\subsubsection{\label{sec:def_delay}Clogging delays}

Clogging delays are a quite relevant quantity during the evacuation process. 
They are defined as the period of time between the egress of two consecutive 
pedestrians \cite{dorso_2007,dorso_2005,dorso_2011}. Animations shows that 
there are two generating mechanisms for a delay. \\

\subsubsection*{\label{sec:gran_delay}Frictional clogging delays}

\begin{figure}[!ht]
\centering
\includegraphics[width=0.5\columnwidth]{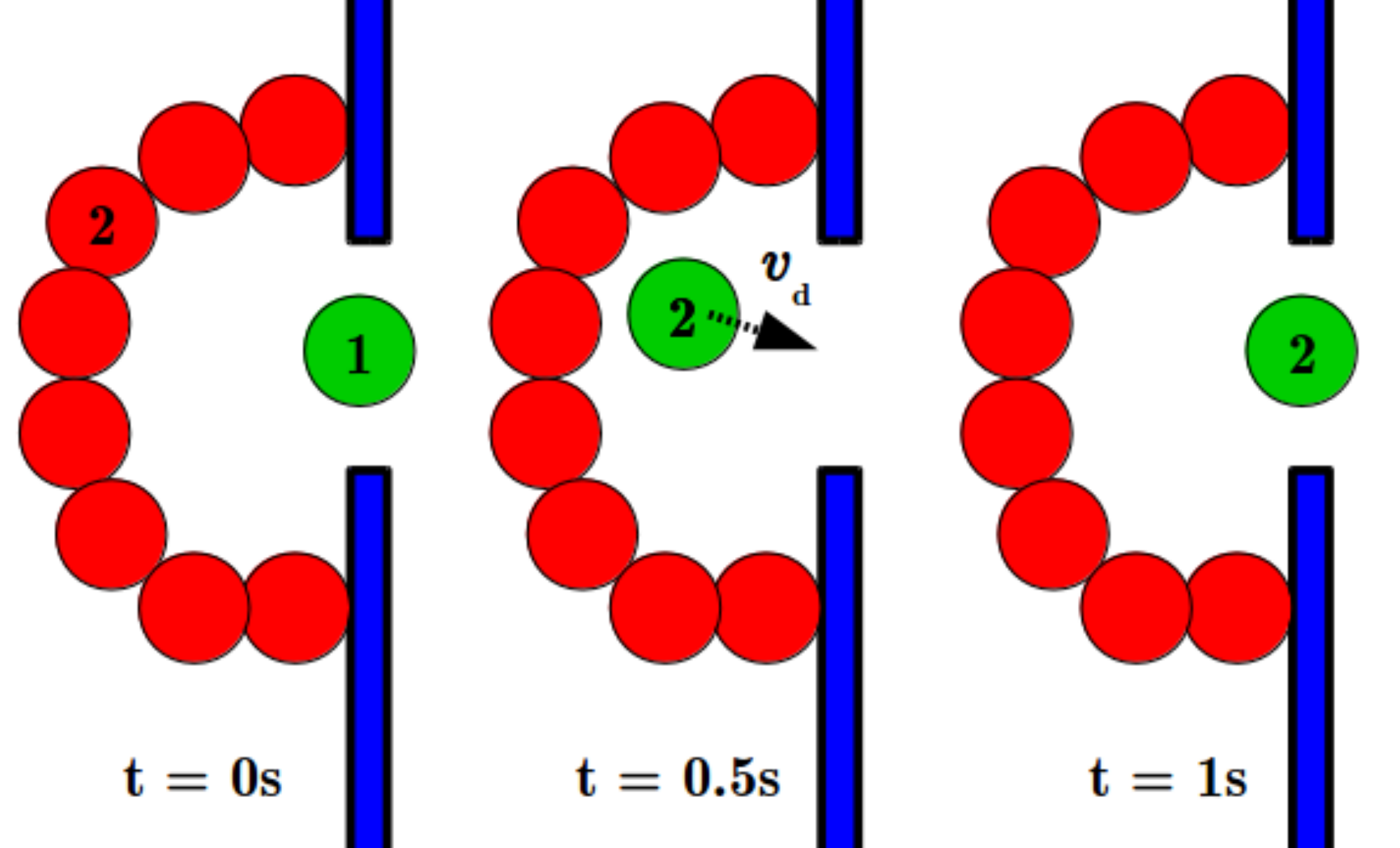}
\caption{\label{fig:mechanism_delay_diag_1} Schematic representation of 
the generating mechanism of a frictional clogging delay. The black line 
corresponds to the direction of the desired velocity $v_d$. A blocking 
cluster (red circles) impedes ind.$\,2$ from leaving the room. It 
breaks at $t=0.5\,$s and finally, at $t=1\,$s he (she) escapes from the room. 
See text for 
more details.}
\end{figure}

As stated in the last Section, blocking clusters are those spatial structures 
that block the exit. Refs.~\cite{dorso_2007,dorso_2005,dorso_2017} show that 
they are responsible of worsening the evacuation performance. Thus, blocking 
clusters are one of the mechanism of production of a delay. In 
Fig.~\ref{fig:mechanism_delay_diag_1} 
we illustrate how this mechanism operates. Notice that the clogging delay 
starts when individual ``1'' left the room. At the same time (\textit{i.e.}, 
$t=0\,$s), we can observe that individual ``2'' belongs to a blocking 
cluster, delaying his (her) outgoing. Later, at $t=0.5\,$s, it breaks and 
finally, individual ``2'' leaves the room at $1\,$s. \\

We conclude that the clogging delay of $1\,$s between individuals 1 and 2 
was originated through a blocking cluster. In other words, the delay is a 
consequence of the granular forces between pedestrians. So, we define a 
\textit{frictional clogging delay} as the clogging delays produce by a blocking 
cluster. \\

We remark that the frictional clogging delay consists of two contributions. One 
corresponds to the time that spends the individual jammed in the blocking 
cluster. And the other corresponds to the time lapse between the breakup of the 
blocking structure and the exit time of the pedestrians (belonging to this 
blocking structure). This time corresponds to the \textit{transit time} 
($0.5\,$s in Fig.~\ref{fig:mechanism_delay_diag_1}). \\

\subsubsection*{\label{sec:social_delay}Social clogging delays}

\begin{figure*}[!ht]
\subfloat[\label{fig:mechanism_delay_diag_2}]{
\includegraphics[scale=0.28]{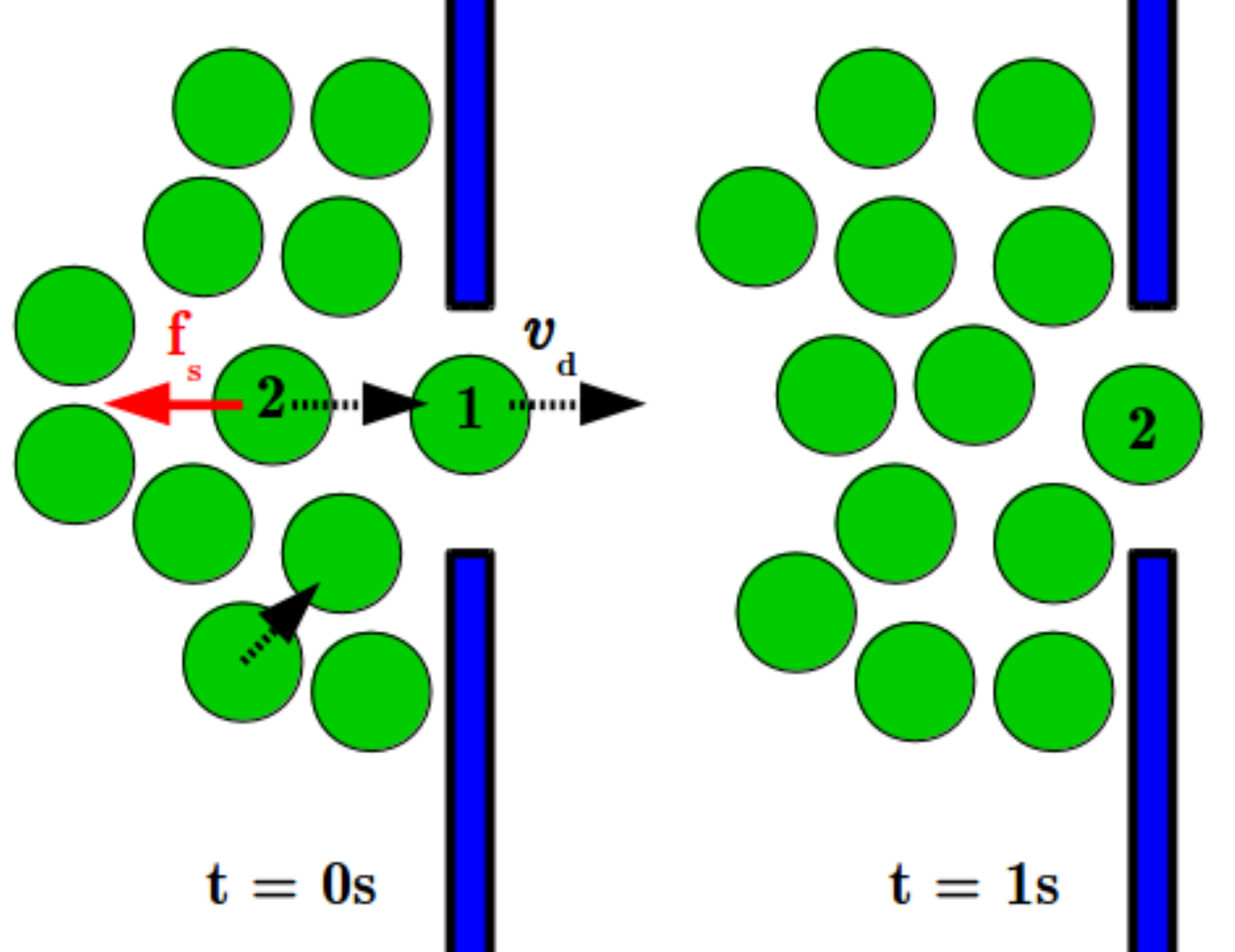}
}
\hspace{2mm}
\subfloat[\label{fig:mechanism_delay_diag_3}]{
\includegraphics[scale=0.3]{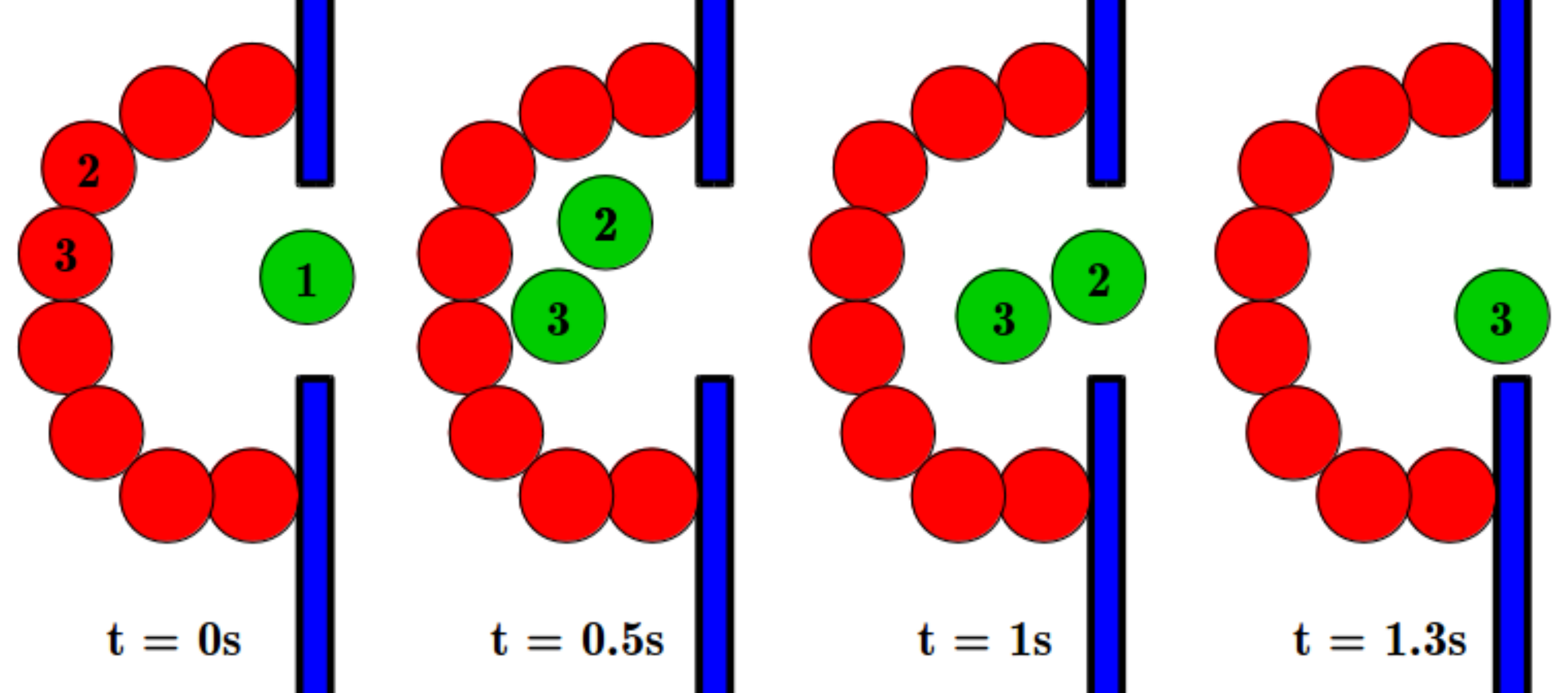}
}
\caption{\label{fig:mechanism_social_delay} (a,b) Schematic representation 
of the production mechanism of a social clogging delay. Black lines correspond 
to the direction of the desired velocity $v_d$. The red continue line 
corresponds to the social force $f_s$ exerted by individual ``1'' over ``2''. 
Green (red) circles indicates that this pedestrian is (is not) in contact with 
his (her) surrounding individuals. (a) At $t=0\,$s, pedestrians cannot leave 
the room due to the presence of individual 1. When he (she) escapes, 
pedestrian labeled as 2 leaves the room at $1\,$s. (b) A blocking cluster (red 
circles) impedes individual 2 and 3 (among others) to leave the room. It breaks 
at $t=0.5\,$s and later, at $t=1\,$s, individual 2 escapes from the room. 
Finally, pedestrian 3 escapes at $t=1.3\,$s. The social clogging delay 
corresponds to the time lapse between individual 2 and 3 (\textit{i.e.} $\Delta 
t=0.3\,$s). Again, delay of $1\,$s between pedestrians 1 and 2 corresponds to a 
frictional clogging delay.} 
\end{figure*}

Animations show that there exists another generating mechanism for a delay. 
On one side, this can occur without the presence of a blocking 
cluster, as we illustrate in Fig.~\ref{fig:mechanism_delay_diag_2}. As in 
Fig.~\ref{fig:mechanism_delay_diag_1}, the delay of $1\,$s corresponds to the 
outgoing process between individuals 1 and 2. But, in this case, 
he (she) is delayed due to the presence of individual 1. Notice that this 
occurs due to the social force. \\

We stress the fact that this situations occur for low pedestrian 
density. This may be the case when the anxiety level is small and therefore, 
individuals do not each other. Thus, we can observe an unstable equilibrium 
between the desired force and the social force near the door. \\

On other side, we have the situation in which the social force delayed the 
outgoing of people can be reached through the breaking of a blocking cluster. We 
illustrate this mechanism in Fig.~\ref{fig:mechanism_delay_diag_3}. Again, a 
blocking cluster impedes many pedestrians to leave the room, particularly 
individuals labeled as 2 ans 3. But, unlike 
Fig.~\ref{fig:mechanism_delay_diag_1}, it releases two pedestrians after it 
breaks, like a \textit{burst}, at $t=0.5\,$s. Notice that $t=1\,$s 
corresponds to the same situation showed in 
Fig.~\ref{fig:mechanism_delay_diag_2} at $t=0\,$s. But, in this case the 
clogging delay of $0.3\,$s between 2 and 3 is shorter than the former 
(\textit{i.e.}, $\Delta t=1\,$s). Finally, notice that the clogging delay of 
1~s between individuals 1 and 2 corresponds to a frictional clogging 
delay (\textit{i.e.}, produced by a blocking cluster). \\

Summarizing, we conclude that the clogging delay of $1\,$s and $0.3\,$s in 
Fig.~\ref{fig:mechanism_delay_diag_2} and 
Fig.~\ref{fig:mechanism_delay_diag_3}, respectively, was produced by the 
repulsion between pedestrians. In other words, the delay is a consequence 
of the social force between them. So, we define a \textit{social 
clogging delay} as those clogging delays produced by the social interaction. \\

\subsubsection{\label{sec:cac}Arch-clogging correlation coefficient}

In order to quantify the relationship between blocking cluster and clogging 
delay we define the ‘‘arch-clogging’’ correlation coefficient as follows 
\cite{dorso_2005} 

\begin{equation}
 c_{ac} = \frac{1}{N}\sum_{cd=1}^{N}f(t_2^{bc},t_1^{cd},t_2^{cd}), 
\label{eq:corr}
\end{equation}

\noindent where $N$ is the total number of clogging delays during the reported 
time interval, $t_2^{bc}$ corresponds to the time stamp for the blocking 
cluster breaking and $t_1^{cd}$ ($t_2^{cd}$) corresponds to the time stamp for 
the delay beginning (ending). The function $f$ is equal to 1 if $t_1^{cd}\leq 
t_2^{bc}\leq t_2^{cd}$, and zero otherwise. This means that $f$ is equal to 
1 if the outgoing individual belonged to a blocking cluster during the interval 
$(t_1^{cd},t_2^{cd})$. Table~\ref{tab:coef_delays} resumes the diagrams shown 
in Figs.~\ref{fig:mechanism_delay_diag_1} and 
\ref{fig:mechanism_social_delay}.\\

\begin{table}
\caption{Computation of the arch-clogging correlation coefficient 
$c_{ac}$ (Eq.~\ref{eq:corr}) in the case of 
Figs.~\ref{fig:mechanism_delay_diag_1} and 
\ref{fig:mechanism_social_delay}.}
\centering 
\begin{center}
\begin{tabular}{ccccccc@{\hspace{4mm}}c@{\hspace{1mm}}l}
 \hline
 Fig. & Pairwise (i,j) & $t_1^{cd}$~(s) & $t_2^{bc}$~(s) & $t_2^{cd}$~(s) & 
$f^{(i,j)}$ & $\Delta t$~(s) & clogging type\\
 \hline  
\ref{fig:mechanism_delay_diag_1}  & (1,2) & 0 & 0.5 & 1 & 1 & 1 & frictional \\
\ref{fig:mechanism_delay_diag_2}  & (1,2) & 0 & - & 1 & 0 & 1 & social \\
\ref{fig:mechanism_delay_diag_3}  & (1,2) & 0 & 0.5 & 1 & 1 & 1 & frictional \\
\ref{fig:mechanism_delay_diag_3}  & (2,3) & 1 & - & 1.3 & 0 & 0.3 & social \\
\hline
\end{tabular}\label{tab:coef_delays}
\end{center}

\end{table}

This correlation $c_{ac}$ represents the fraction of \textit{frictional 
clogging delays} (or \textit{social clogging delays}) with respect to all the 
clogging delays appearing during the (stationary) evacuation process. Any value 
close to one indicates that most of the clogging delays belong to 
blocking clusters. On the contrary, if $c_{ac}$ is close to zero, most 
of them belong to social force interactions.

\section{\label{sec:simulations}Numerical simulations}

The simulations were performed on a $20\,$m$~\times$ $20\,$m square room with 
225 pedestrians inside. The occupancy density was set to $0.6\,$people/m$^2$, as 
suggested by healthy indoor environmental regulations \cite{mysen}. The room had 
a single exit on one side, placed in the middle of it in order to avoid corner 
effects. The door width was $L=0.96\,$m, enough to allow up to two pedestrians 
to escape simultaneously (side by side). \\

The pedestrians were modeled as soft spheres. They were initially 
placed in a regular square arrangement  along the room with random velocities, 
resembling a Gaussian distribution with null mean value. The rms value for the 
Gaussian distribution was close to $1\,$m/s. The desired velocity $v_d$ was the 
same for all the individuals. At each time-step, however, the desired direction 
$\mathbf{e}_d$ was updated, in order to point to the exit. \\

We used periodic boundary condition (re-entering mechanism) for the outgoing 
pedestrians. That is, those individuals who were able to leave the room 
were reinserted at the back end of the room and placed at the very back of the 
bulk with velocity v~=~0.1$\,$m/s, in order to cause a minimal bulk 
perturbation. This mechanism was carried out in order to keep the crowd size 
unchanged, and therefore, the pressure among pedestrians. \\

According to the literature (see Ref~\cite{helbing_2000}), the model 
parameters used were $\tau=0.5\,$s, $A=2000\,$N, $B=0.08\,$m and 
$\kappa_t=2.4\times$~10$^5\,$kg~m$^{-1}~s^{-1}$. However, the 
pedestrian's mass and radius were set according to the more realist values of 
$70\,$kg and $0.23\,$m, as in Ref.~\cite{Sticco_2020}. Also, the 
compression coefficient $\kappa_n$ was set equal to 
2.62~$\times$~10$^4$~N~m$^{-1}$, according to the experimental value of the 
human torso stiffness \cite{Cornes1}. \\

The simulations were performed using {\sc Lammps} molecular dynamics simulator 
with parallel computing capabilities \cite{plimpton}. The time integration 
algorithm followed the velocity Verlet scheme with a time step of 
$10^{-4}\,$s. We implemented special modules in C++ for upgrading the {\sc 
Lammps} capabilities to attain the ``social force model'' simulations. We also 
checked over the {\sc Lammps} output with previous computations (see 
Refs.~\cite{dorso_2007,dorso_2005}).\\

Data recording was done at time intervals of 0.05$\,\tau$, that is, at 
intervals as short as 10\% of the pedestrian's relaxation time. The simulating 
process lasted until 7000 pedestrians left the room. \\ 

The explored anxiety levels ranged from relaxed situations ($v_d=1\,$m/s) 
to highly stressing ones ($v_d=10\,$m/s). This last value can hardly be reached 
in real life situations. However, in Ref.~\cite{dorso_2017}, it was shown that 
similar pressure levels can be reached in a big crowd with a moderate anxiety 
level. Thus, this wide range of desired velocities allows us to study different 
pressure scenarios. We also assumed that the pedestrians were not able to fall 
due to the crowd pressure as in Ref.~\cite{Cornes1}. \\

\section{\label{sec:results}Results}

Our results run along four major streams as follows:

\begin{itemize}
 \item In Section~\ref{sec:revision} we revisit the context of the evacuation 
time. We further introduce other novel concepts that will be used throughout the 
work. 

 \item In Section~\ref{sec:delays} we show the differences between the 
\textit{faster is slower} and \textit{faster is faster} regimes in terms of the 
clogging delays.

\item In Section~\ref{sec:clusters} we focus on the clusterization phenomenon 
at either the \textit{faster is slower} and \textit{faster is faster} regimes.

 \item In Section~\ref{sec:relation_gran_delays} we analyze the relationship 
between the size of the clustering structures and the corresponding clogging 
delays. \\
\end{itemize}

\subsection{\label{sec:revision}A review of the evacuation processes at 
bottlenecks}

\subsection*{\label{sec:fis}The evacuation time versus the desired velocity}

\begin{figure}[!ht]
\centering
\includegraphics[width=0.7\columnwidth]{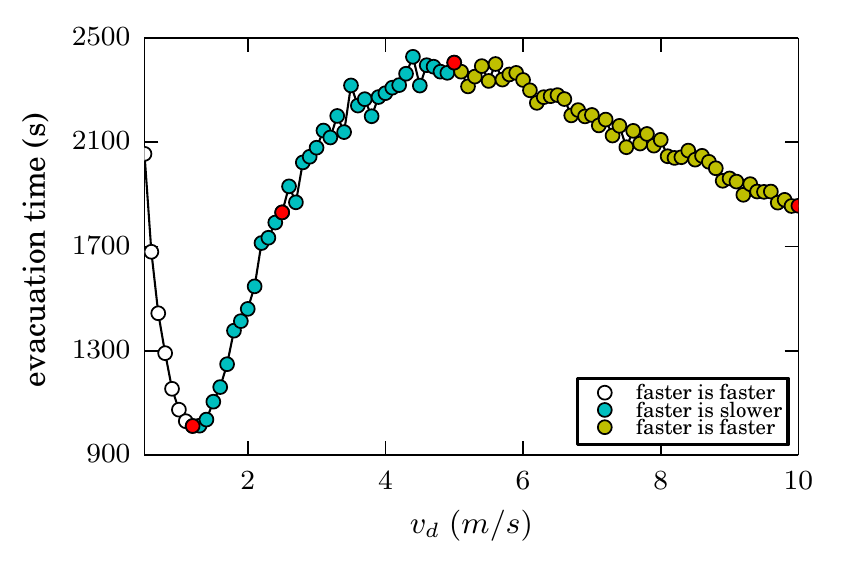}
\caption{\label{fig:fis} Evacuation time for the first 7000 evacuees as a 
function of the desired velocity $v_d$. The simulated room was $20\,$m$~\times$ 
$20\,$m with a single door of $0.92\,$m (say, the diameter of two pedestrians). 
The number of individuals inside the room was 225. This curve was computed from 
a single simulation process. Re-entering mechanism was allowed (see text for 
more details). The simulation lasted until 7000 individuals left the room. 
Desired velocities of $v_d$~=~1.2, 2.5, 5.0 and 10.0$\,$m/s are indicated in red 
color.}
\end{figure}

As a first step, we computed the evacuation time for a wide range of desired 
velocities $v_d$. This is shown in Fig.~\ref{fig:fis}. We stress the fact that 
the explored range corresponds to the one analyzed by Sticco \textit{et. al.} 
(see Ref.~\cite{dorso_2017}) but, considering periodic boundary conditions. 
This means that any pedestrian who left the room is introduced on the opposite 
side of the bottleneck (see Section~\ref{sec:simulations}). \\

It can be seen in Fig.~\ref{fig:fis} the \textit{faster is slower} (blue 
circles) and the \textit{faster is faster} (white and yellow circles) phenomena 
as reported in Ref.~\cite{dorso_2017}. Recall that the \textit{faster is slower} 
regime, introduced by Helbing \textit{et. al.} \cite{helbing_2000} corresponds 
to the increase in the evacuation time as the pedestrian's anxiety level 
increases. Thus, the evacuation efficiency reduces within this region. It was 
shown in Ref.~\cite{dorso_2007} that this is related to the presence of sliding 
friction among pedestrians (and the walls). \\

Besides, it was reported in Ref.~\cite{dorso_2017} that the evacuation time 
decreases for very high desired velocities (\textit{i.e.}, high pressures). 
This behavior corresponds to the \textit{faster is faster} regime. It was 
concluded (see Ref.~\cite{dorso_2017}) that as the crowd pushing force 
increases, the sliding friction at the ``blocking cluster'' seems not enough to 
prevent this kind of structure from (completely) stopping the crowd movement. 
This is the reason for the improvement of the evacuation time at the exit. We 
call the attention that we are not considering the possibility of fallen 
pedestrians due to high pressures (see Ref.~\cite{Cornes1} for details). 
Further readings on the role of the sliding friction can be found in 
Refs.~\cite{dorso_2007,dorso_2005,dorso_2017}. \\
 
As can be noticed from Fig.~\ref{fig:fis}, the \textit{faster is slower} regime 
appears for desired velocities between 1 and $5\,$m/s (approximately). This is 
somewhat shifted from the range reported in Ref.~\cite{dorso_2017} 
($2\,<v_d<8\,$(m/s), approximately). The same occurs for both \textit{faster is 
faster} effects. This discrepancy is a consequence of the periodic boundary 
condition. The \textit{bulk} pressure remains (almost) constant during the 
stationary state for periodic boundary condition. But, it decreases for 
non-periodic conditions, since pedestrians escape from the room. This phenomenon 
is similar the one observed in Ref.~\cite{dorso_2017} when varying the number of 
pedestrians inside the room. \\

We will focus on four specific desired velocities in the next section. 
These correspond to the minimum evacuation time ($v_d=1.2\,$m/s), the 
maximum evacuation time ($v_d=5\,$m/s) and two desired velocities with similar 
evacuation times but on different regimes. The desired velocities $v_d=2.5\,$m/s 
and $v_d=10\,$m/s were chosen as representative of the \textit{faster is slower} 
and \textit{faster is faster} phenomena, respectively.\\

\subsubsection*{\label{sec:descargas}Discharge curves}

\begin{figure}[!ht]
\centering
\includegraphics[width=0.7\columnwidth]{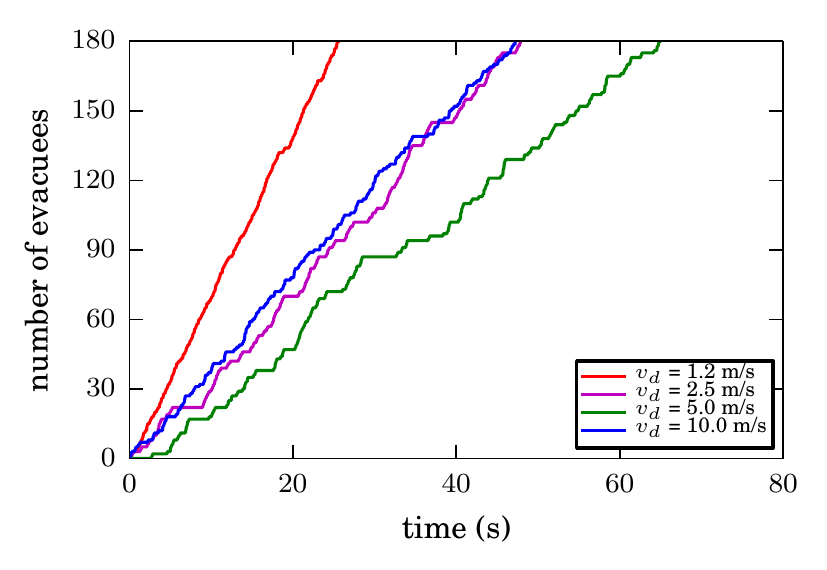}
\caption{\label{fig:discharge} Number of pedestrians that left the 
room versus time, for four different desired velocities $v_d$ (see legend). 
The showed time-window corresponds to a representative sample of 180 
pedestrians that left the room from a set of 7000 evacuees. The analyzed 
desired velocities are indicated by red circles in Fig.~\ref{fig:fis}. The 
curves were computed from a single simulation process (re-entering mechanism was 
allowed)}
\end{figure}

We proceed to a first microscopic analysis of the evacuation process through the 
discharge curves (say, the number of individuals that escape from the room over 
time). Fig.~\ref{fig:discharge} shows the discharge curves for four 
desires velocities (see caption for details) 
\cite{dorso_2005,dorso_2007,dorso_2017}. For each curve, the time 
between consecutive evacuees (\textit{i.e.} the delay) is represented by a 
horizontal line.  \\

It can be seen in Fig.~\ref{fig:discharge} that the evacuation time for the 
first 180 evacuees corresponds to the upper end of each discharge curve. 
It can be seen that the desired velocities $v_d=1.2$ and $5\,$m/s correspond to 
the minimum and maximum evacuation time, respectively (say, $25\,$s and $65\,$s, 
approximately). The curves for $v_d=2.5$ and $10\,$m/s meet at the top of the 
figure, as expected from Fig.~\ref{fig:fis}. But most interesting, it can be 
noticed an increase in the duration of the delays (represented by horizontal 
segments) from the curve on the left to one on the right in 
Fig.~\ref{fig:discharge}. It is worth mentioning that the evacuation time 
corresponds to the sum of all delays, and thus, the increase in evacuation time 
is related to the increase in the duration of delays. The relationship between 
evacuation time and delays will be further analyzed in more detail in the next 
section. \\

A statistical test on the (global) uniformity of the discharge curves was done 
in \ref{sec:K-S}. It could be established that the flow of evacuees is not 
uniform at the \textit{faster is slower} and \textit{faster is faster} regimes 
(within a significance of $5\%$). However, a tendency towards uniformity could 
be noticed (at least) for the \textit{faster is faster} regime as the 
individual's anxiety level ($v_d$) increases. \\

In summary, the discharge curves show that the pedestrian flow is never 
uniform (within a significance level), but delays tends to become more regular 
for very high anxiety levels (or \textit{bulk} pressures). \\

\subsection{\label{sec:delays}Clogging delays analysis}

In this section we turn to study the mechanism by which different types of 
clogging delays are generated. We analyze the clogging delays distributions and 
the contribution of each type of delay to the overall evacuation time. \\

\subsubsection{\label{sec:corr_coef}Production mechanism of a delay}

\begin{figure}[!ht]
\centering
\includegraphics[width=0.7\columnwidth]{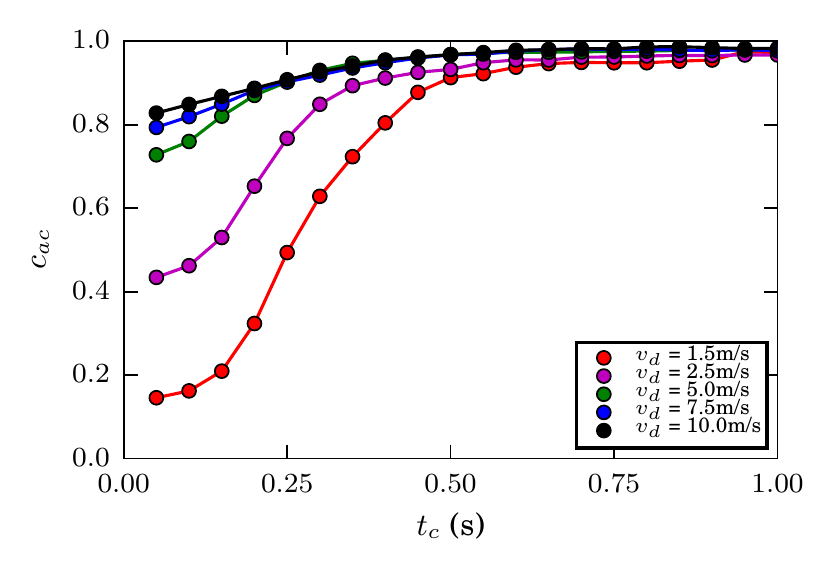}
\caption{\label{fig:variando_tc} Arch-clogging correlation coefficient $c_{ac}$ 
as a function of the duration of the delay (equal or greater than $t_c$) for 
five different desired velocities (see legend for details). All the curves were 
computed from a single simulation process (re-entering mechanism was allowed). 
We computed 6999 clogging delays for each desired velocity, according to the 
evacuation of 7000 pedestrians. }
\end{figure}

As mentioned in Section~\ref{sec:def_delay}, there are two categories of 
clogging delays. The first one corresponds to those generated as a 
consequence of the social force among individuals (see 
Fig.~\ref{fig:mechanism_social_delay}). The second one corresponds to those 
that occur due to blocking clusters (see Fig.~\ref{fig:mechanism_delay_diag_1}). 
The former corresponds to a \textit{social clogging delay}, while the latter 
corresponds for a \textit{frictional clogging delay}. Fig.~\ref{fig:variando_tc} 
plots the coefficient $c_{ac}$ (see Eq.~\ref{eq:corr}) as a function of the 
clogging delays longer than (or equal to) any threshold $t_c$.\\

According to Fig.~\ref{fig:variando_tc}, the probability of finding frictional 
delays increases among long delays. In other words, long lasting delays 
commonly belong to blocking clusters. Thus, we can (almost) exclusively 
classify any delay longer than $1\,$s as a \textit{frictional clogging delay}. 
Notice that this criterion is fulfilled regardless of the value of the 
desired velocity $v_d$ (within the explored range).  The situation for short 
clogging delays appears somewhat mixed. Not all delays less than $1\,$s belong 
to a blocking cluster. \\

Furthermore, the fraction of frictional delays increases for increasing desired 
velocities (see Fig.~\ref{fig:mechanism_social_delay}). Recall from 
\ref{sec:prob_bc} that the probability of blocking clusters also increases for 
increasing values of $v_d$. This is the reason for the increase of the fraction 
of frictional clogging delays when increasing the desired velocity.\\


A careful examination of the evacuation animations shows that social delays may 
also appear after the breaking process of a blocking cluster. This corresponds 
to the \textit{burst} released after the rupture, as exhibit in 
Fig.~\ref{fig:mechanism_delay_diag_3}. We observed in the animations (not 
shown) that this phenomenon occurs commonly as the desired velocity increases. 
Therefore, social delays should not be considered as ``opposed'' to frictional 
delays, but complementary to these. \\

The major conclusion from this Section is that long delays (say, greater than 
$1\,$s) can be associated to blocking clusters. But, those delays of short 
duration may either be associated to social interactions or granular 
interactions. \\

\begin{figure}[!ht]
\centering
\includegraphics[width=0.7\columnwidth]{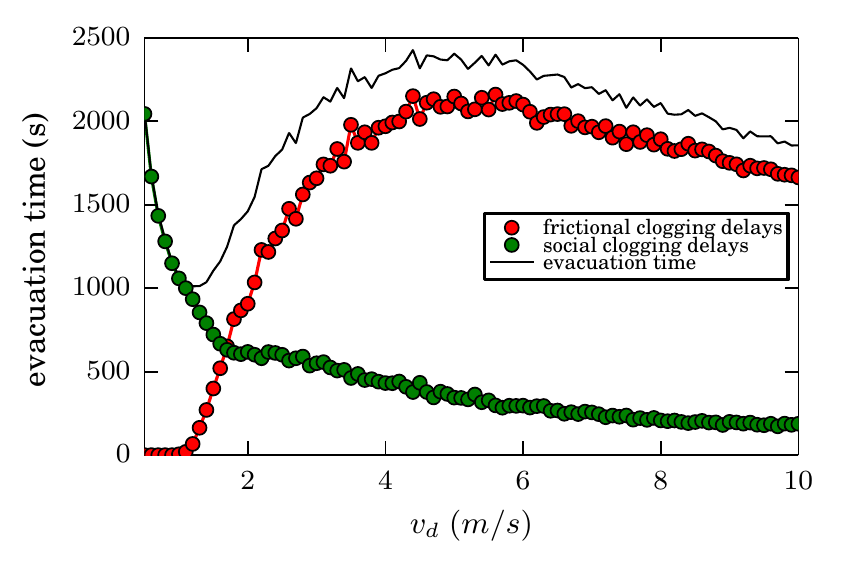}
\caption{\label{fig:tiempo_bcs} Evacuation time for the first 7000 evacuees as 
a function of the desired velocity $v_d$. The black line corresponds to the 
total evacuation time. Red and green circles correspond to the evacuation time 
associated with frictional and social clogging delays, respectively. These were 
recorded from a single simulation process (re-entering mechanism was 
allowed).}
\end{figure}

Our next step focuses on the relevance of the frictional and social clogging 
delays on the evacuation time. We computed the evacuation time considering 
these two categories of delays separately. Fig.~\ref{fig:tiempo_bcs} shows the 
results. \\

The evacuation time considering (adding) only the frictional delays (and thus 
the blocking clusters) is qualitatively similar to the total evacuation time. 
This corresponds to (almost) the entire explored interval ($v_d>1.2\,$m/s). 
However, this vanishes for $v_d~<~1.2\,$m/s. Recall that people are very seldom 
in contact between each other at such low desired velocities, and therefore, no 
blocking clusters are present (see \ref{sec:prob_bc}). \\

The evacuation time associated to the social delays (\textit{i.e.} social 
clogging delays) decreases monotonically as the desired velocity increases. As 
mentioned above, the individuals are very seldom in contact 
for $v_d<1.2\,$m/s (see Fig.~\ref{fig:mechanism_delay_diag_2}). Notice that 
the total evacuation time and the social evacuation time coincide for 
$v_d<1.2\,$m/s, but no \textit{faster is slower} effect takes place for the 
latter at higher $v_s$'s (as expected). \\

\subsubsection{\label{sec:distribuciones}Clogging delay distributions}

\begin{figure*}[!ht]
\subfloat[$v_d$~=~1.2~m/s (minimum $t_{ev}$)\label{fig:histo_delays_vd_1_2}]{
\includegraphics[scale=0.815]{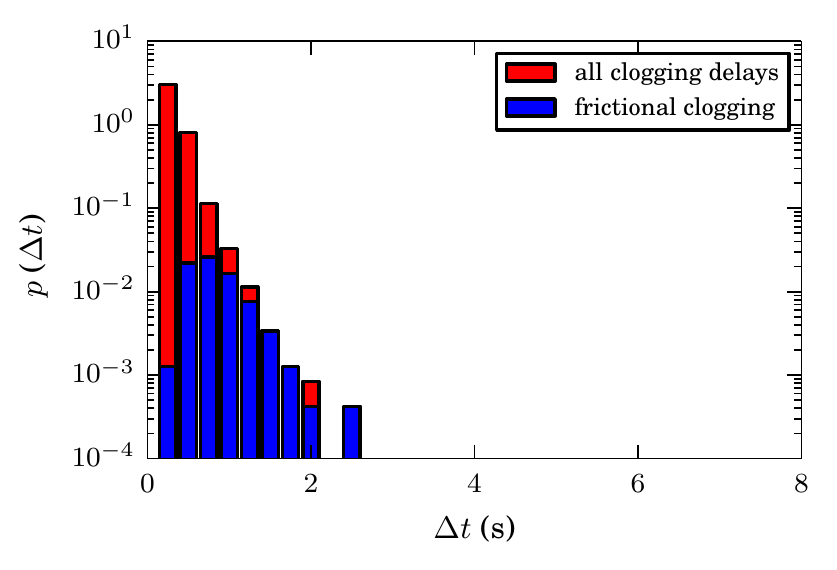}
}
\hspace{-2mm}
\subfloat[$v_d$~=~2.5~m/s\label{fig:histo_delays_vd_2_5}]{
\includegraphics[scale=0.815]{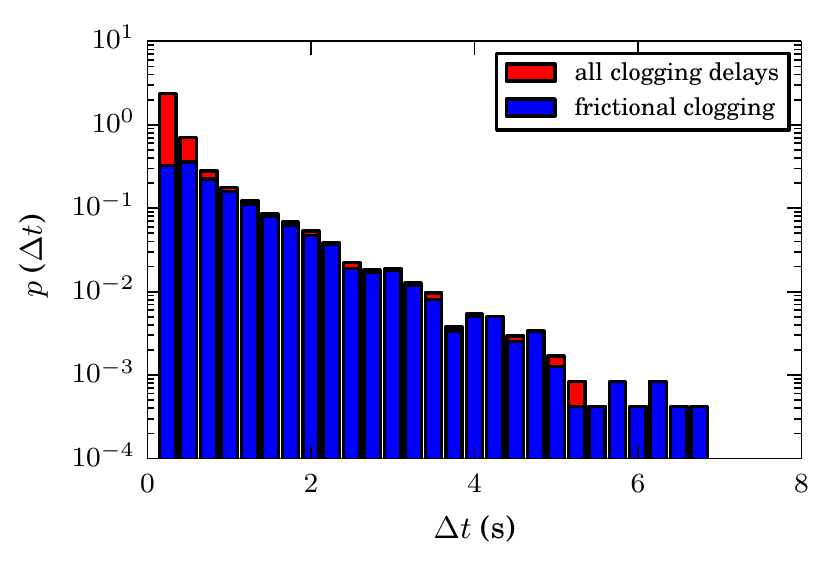}
}
\hspace{-2.5mm}
\subfloat[$v_d$~=~5~m/s (maximum $t_{ev}$)\label{fig:histo_delays_vd_5}]{
\includegraphics[scale=0.815]{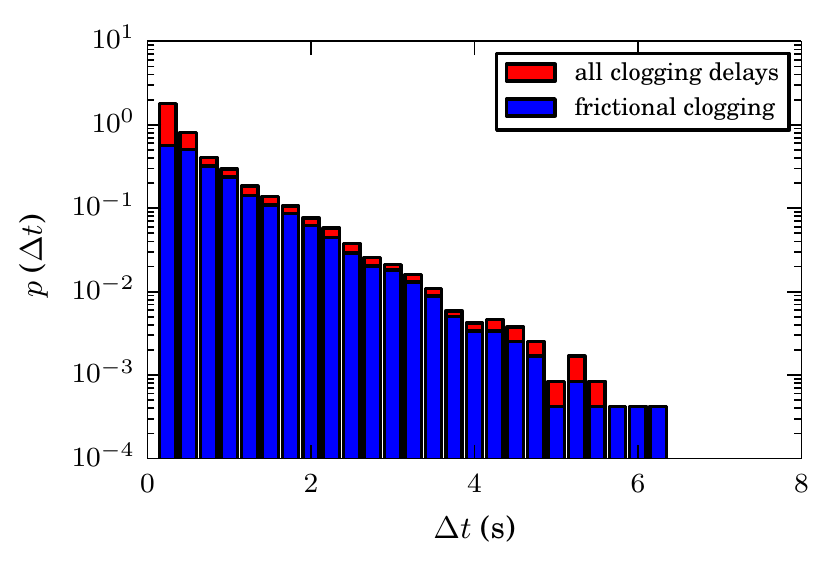}
}
\hspace{-2.5mm}
\subfloat[$v_d$~=~10~m/s\label{fig:histo_delays_vd_10}]{
\includegraphics[scale=0.815]{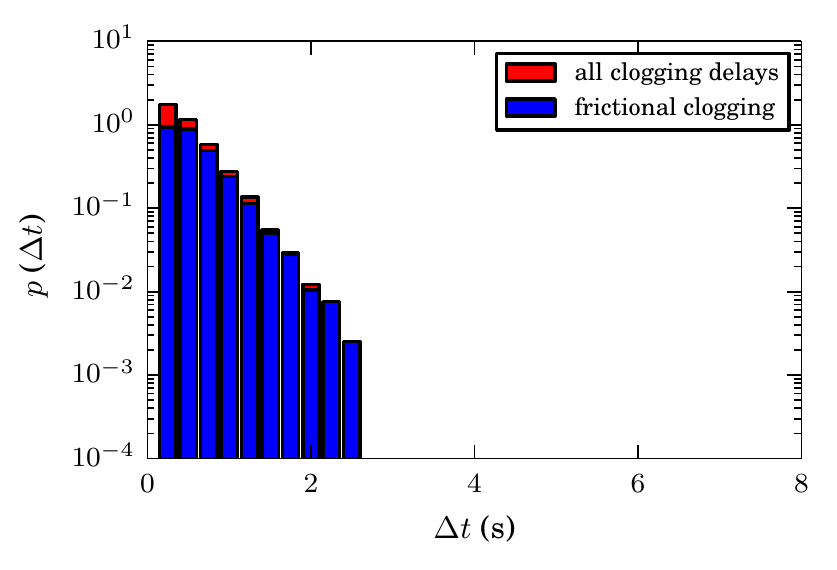}
}
\caption{\label{fig:histo} Normalized distributions of time lapses $\Delta t$ 
between egresses of consecutive pedestrians up to 7000 people evacuated, for 
the same values of $v_d$ analyzed in Fig.~\ref{fig:discharge}. The bin size is 
$0.25\,$s. Clogging delays and delays originated due to a blocking cluster 
(named frictional clogging) are indicated by red and blue bars, 
respectively. Frictional clogging corresponds to a subset of the clogging 
delays (see text for details). Thus, red bars are plotted behind the blue bars. 
The distribution area corresponding to the clogging delays is 1. Data was 
recorded after the first $10\,$s of the beginning of the simulation process, in 
order to avoid non-stationary effects. The door width was $0.92\,$m (say, the 
diameter of two pedestrians).} 
\end{figure*}

We further computed the delays's probability distribution, as shown in 
Fig.~\ref{fig:histo}. We chose the same desired velocities as in 
Section~\ref{sec:descargas} for the purpose of comparison. The distribution 
corresponding to all the clogging delays and the \textit{frictional clogging 
delays} can be seen in red and blue bars, respectively. The latter corresponds 
to a subset of the former. Notice that distributions are valid for the specific 
door width mentioned in the caption (see Ref.~\cite{dorso_2005} for more 
details). \\

The delays shorter than $3\,$s in Fig.~\ref{fig:histo} (approximately) are 
present in all the plotted situations. But, clogging delays greater than $3\,$s 
occur only for the intermediate situations $v_d~=~$2.5 and 5$\,$m/s (see 
Figs.~\ref{fig:histo_delays_vd_2_5} and \ref{fig:histo_delays_vd_5}). Both 
situations exhibit delays up to 7$\,$s approximately. Recall that these levels 
of $v_d$ correspond to the \textit{faster is slower} regime (see 
Fig.~\ref{fig:fis}). The \textit{faster is slower} regime includes long delays, 
while the \textit{faster is faster} regime stands for delays no longer than 
$3\,$s (under the explored conditions). \\

\subsubsection{\label{sec:delays_bineado}Relevance of the clogging delays}

\begin{figure*}[!ht]
\subfloat[Short delays\label{fig:delays_bin_cortos}]{
\includegraphics[scale=0.775]{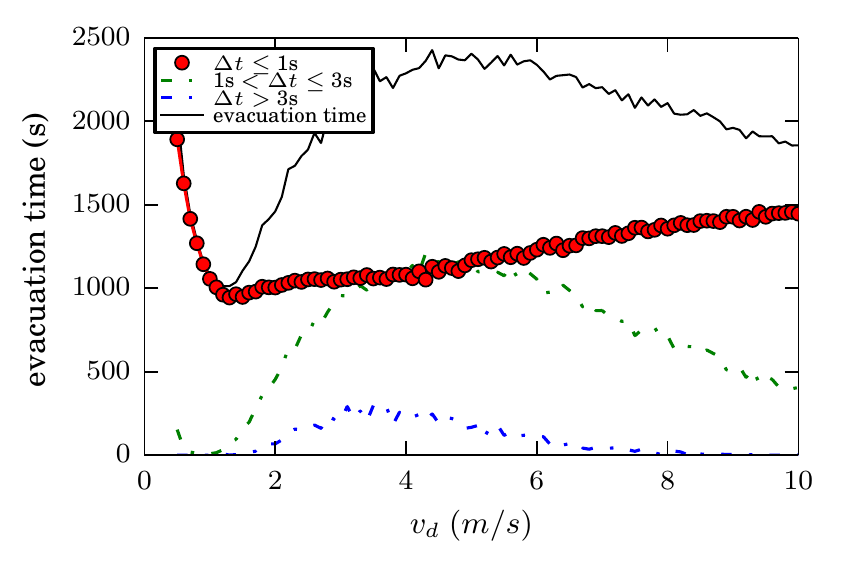}
}
\hspace{-2mm}
\subfloat[Intermediate and long delays\label{fig:delays_bin_largos}]{
\includegraphics[scale=0.775]{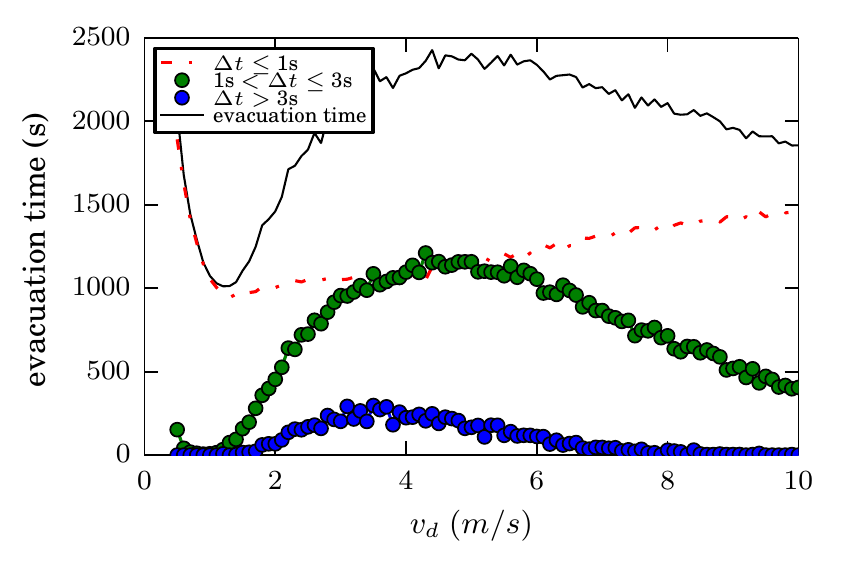}
}
\caption{\label{fig:bineado_delays} The black line corresponds to the 
evacuation time for the first 7000 evacuees as a function of the desired 
velocity $v_d$. Red, green and blue circles (and dashed lines) correspond to 
the different delay categories. } 
\end{figure*}

We concluded in Section~\ref{sec:corr_coef} that almost all the delays longer 
than $1\,$s correspond to the presence of a blocking cluster. Therefore, they 
correspond to \textit{frictional clogging delays}. In 
Section~\ref{sec:distribuciones} we further noticed that long delays 
(greater than $3\,$s) appear during the \textit{faster is slower} regime. We 
focus on three categories: short ($\Delta t \leq1\,$s), intermediate 
(1$\,$s~$<\Delta t \leq3\,$s) and long ($\Delta t >3\,$s) delays. 
Fig.~\ref{fig:bineado_delays} shows the evacuation time for these types of 
delays. \\

\begin{itemize}
 \item Short clogging delays ($\Delta t \leq1\,$s): This kind of delays are the 
only relevant ones for desired velocities below $1.2\,$m/s. We already 
mentioned in \ref{sec:prob_bc} that no relevant blocking clusters appear at 
these desired velocities, and thus, all the clogging delays correspond to 
\textit{social clogging delays}. Also, as can be seen this kind of delay does 
not allow us to distinguish the \textit{faster is slower} and \textit{faster 
is faster} regimes (above $1.2\,$m/s).  \\

\item Intermediate clogging delays (1$\,$s~$<\Delta t \leq3\,$s): This category 
becomes relevant for $v_d>1.2\,$m/s. Their qualitative behavior 
for $v_d>1.2\,$m/s is similar to the total evacuation time. Thus, unlike the 
short clogging delays, these are a good observable to identify the 
\textit{faster is slower} and \textit{faster is faster} regimes. Also, the 
cumulative time due to both delays explains almost all of the evacuation time.\\

\item Long clogging delays ($\Delta t >3\,$s): They are only present for 
velocities between 1 and $7\,$m/s (as we mentioned in 
Section~\ref{sec:distribuciones}). As can be seen, these delays are not 
decisive for the presence of the \textit{faster is slower} effect. Also, the 
lack of long clogging delays may be partially responsible for the \textit{faster 
is faster} phenomenon.
\end{itemize}

The following conclusions can be outlined from this Section. The analysis of 
the delays show that the overall evacuation time is mainly due to 
\textit{frictional clogging delays} for $v_d>2\,$m/s. Also, the slope of 
the intermediate clogging delays (1$\,$s~$<\Delta t \leq3\,$s) allows us to 
distinguish between the \textit{faster is slower} and \textit{faster is faster} 
regimes (see Fig.~\ref{fig:bineado_delays}). \\

\subsection{\label{sec:clusters}Clustering structure analysis}

In this section we turn to study the clustering structures. We analyze the 
different kind of spatial clusters and we explore how they change with the 
desired velocity.

\subsubsection{\label{sec:snapshots}The morphology in the bulk of the crowd}

\begin{figure*}[!ht]
\subfloat[$v_d$~=~1.2~m/s\label{fig:snapshots_vd_1_2}]{
\includegraphics[scale=0.315]{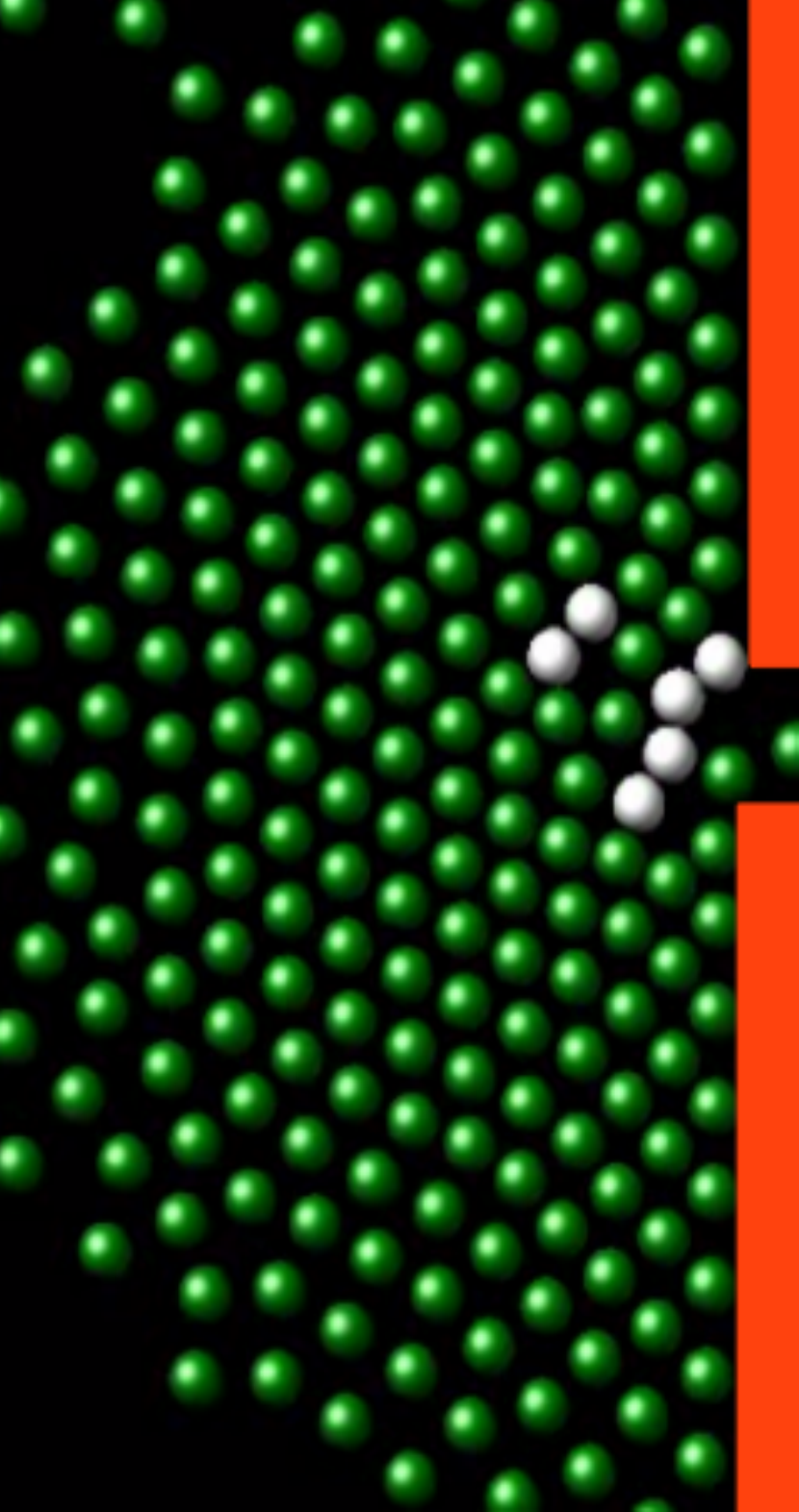}
}
\hspace{3mm}
\subfloat[$v_d$~=~2.5~m/s\label{fig:snapshots_vd_2_5}]{
\includegraphics[scale=0.315]{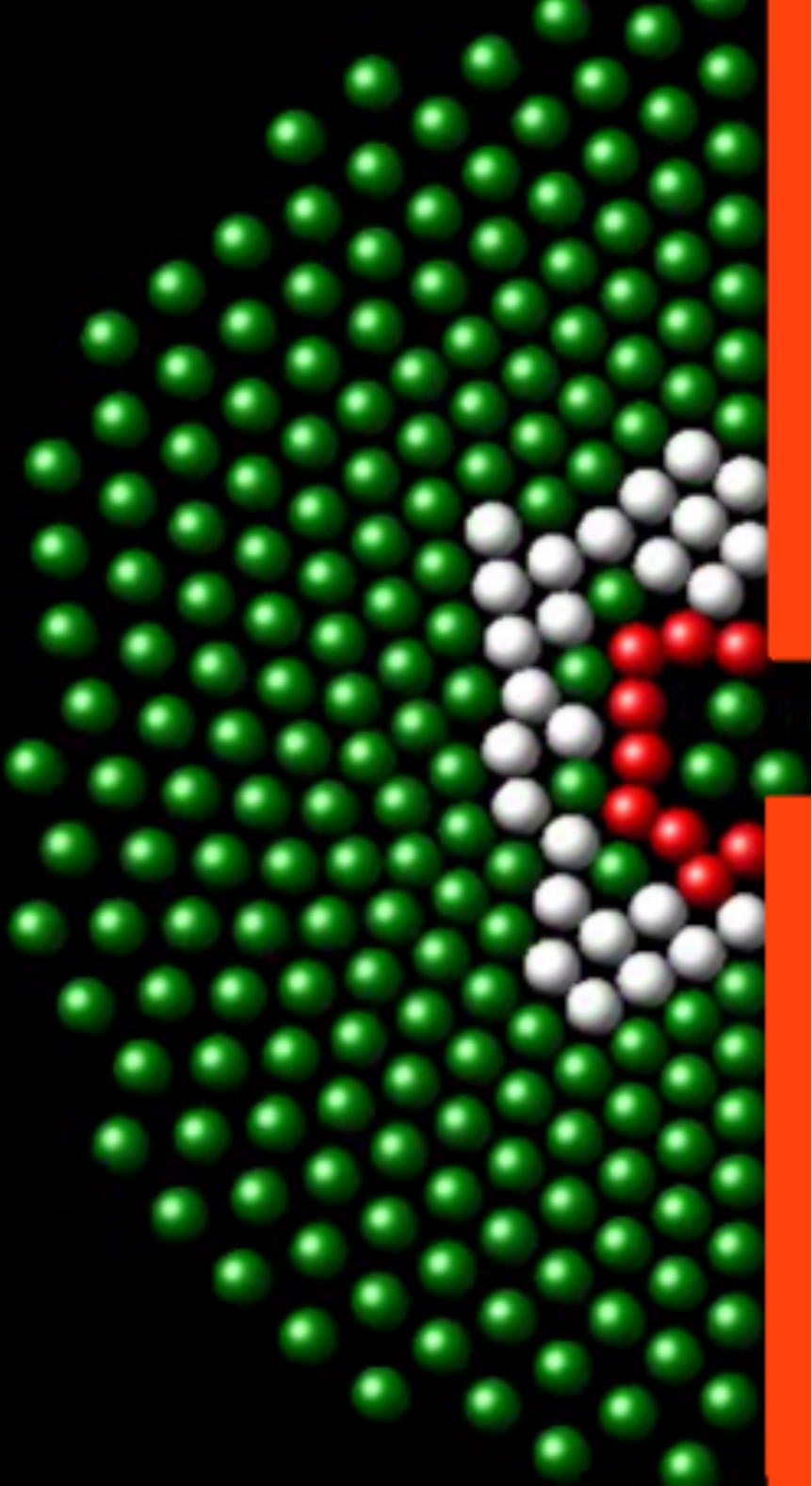}
}
\hspace{3mm}
\subfloat[$v_d$~=~5.0~m/s\label{fig:snapshots_vd_5}]{
\includegraphics[scale=0.335]{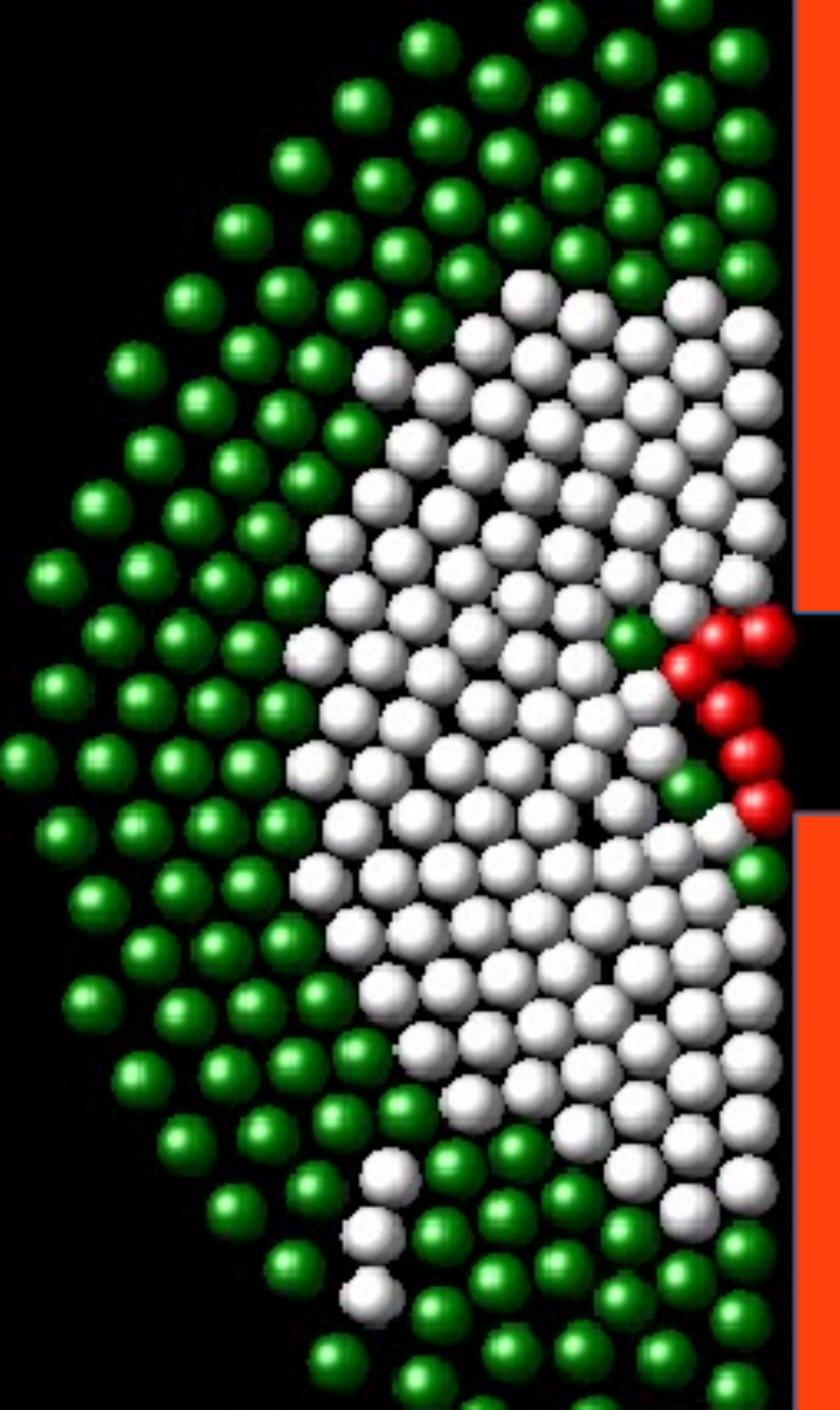}
}
\hspace{3mm}
\subfloat[$v_d$~=~10.0~m/s\label{fig:snapshots_vd_10}]{
\includegraphics[scale=0.315]{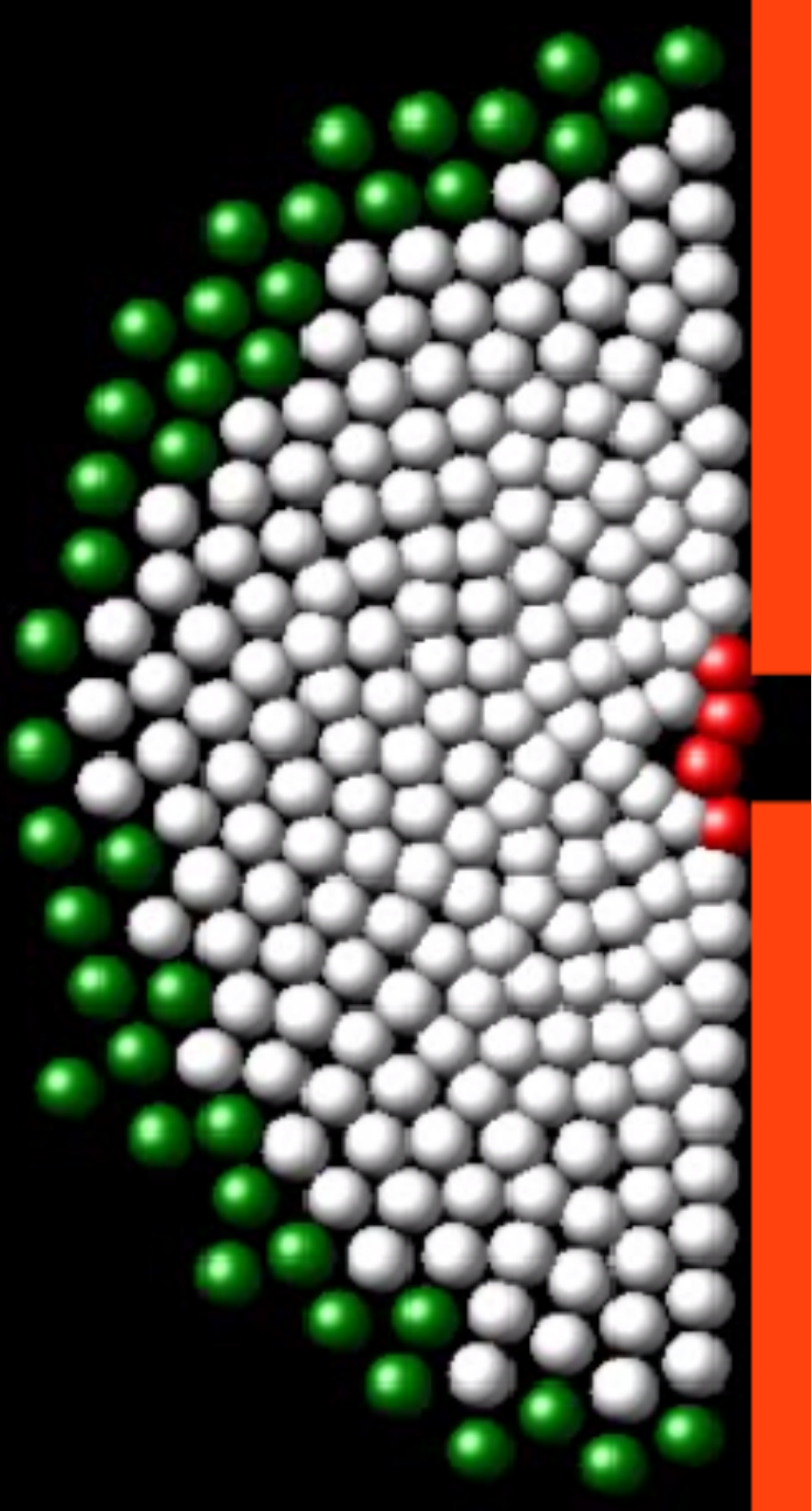}
}
\caption{\label{fig:snapshots} Snapshots of the \textit{bulk} for four desired 
velocities. Red and white circles represent the blocking cluster and the 
spatial clusters, respectively. Green circles correspond to individuals 
without contact with his (her) surrounding pedestrians and walls. The orange 
lines represent the walls on the right of the room. The simulated room was 
$20\,$m$~\times$ $20\,$m with a single door of $0.92\,$m (say, the diameter of 
two individuals). The number of individuals inside the room was 225. The 
selected desired velocities are those indicated in red in Fig.~\ref{fig:fis}.} 
\end{figure*}

Fig.~\ref{fig:snapshots} shows four snapshots of the \textit{bulk} close to 
the exit. The shown situations correspond to the same desired velocities as in 
Section~\ref{sec:delays}. Red and white circles represent the blocking and 
spatial clusters, respectively. We stress that the individuals in red (blocking 
cluster) also belong to the spatial cluster (white color) (see 
Fig.~\ref{fig:snapshots}).\\

As can be seen in Fig.~\ref{fig:snapshots}, the number of individuals in 
contact increases with increasing desired velocities. This is the expected 
picture for people pushing towards the exit. The harder they push (increasing 
values of $v_d$), the larger the clogging region. \\

Notice that two small spatial clusters appear at $v_d=1.2\,$m/s (see white 
circles in Fig.~\ref{fig:snapshots_vd_1_2}). But these merge into a single 
spatial cluster of size 35, as the desired velocity increases to 
$v_d=2.5\,$m/s. However, it can be noticed that 4 individuals (green circles 
in Fig.~\ref{fig:snapshots_vd_2_5}) remain out of contact with their closest 
neighbors. This is a rather common phenomenon for low desired velocities, as we 
were able to observe in the animations (not shown). \\

\begin{figure}[!ht]
\centering
\includegraphics[width=0.7\columnwidth]{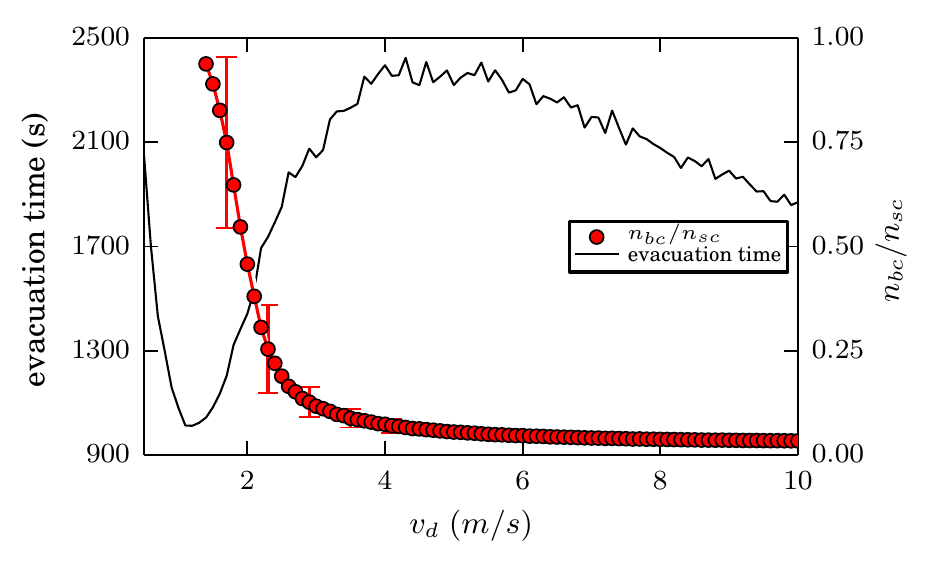}
\caption{\label{fig:size_bc_cluster} The black line corresponds to the 
evacuation time for 7000 evacuees, while the red circles correspond to the 
normalized size of a blocking cluster ($n_{bc}$) with respect to the spatial 
cluster ($n_{sc}$), both as a function of the desired velocity $v_d$. The 
spatial cluster corresponds to those in contact with the blocking cluster. The 
size of the blocking cluster and the spatial cluster were recorded 
every $0.05\,$s. The acquisition was done only when a blocking cluster existed. 
The error bars corresponds to $\pm \sigma$ (one standard deviation). 
These were computed from a single simulation process (re-entering mechanism was 
allowed).}
\end{figure}

Interesting, Fig.~\ref{fig:snapshots_vd_2_5}, also shows a burst of 3 
individuals escaping from the room. These correspond to the breaking of a 
(former) blocking cluster (not shown). Therefore, this is associated to a 
social clogging delay. Also, the snapshots in Figs.~\ref{fig:snapshots_vd_5} 
and \ref{fig:snapshots_vd_10} show a significant change in the size 
relation between the blocking cluster and the spatial cluster for increasing 
desired velocities. The ratio between this magnitudes is shown in 
Fig.~\ref{fig:size_bc_cluster}.\\

A sharp decay in the blocking-to-spatial ratio can be noticed in 
Fig.~\ref{fig:size_bc_cluster}. We immediately identify two regimes according 
to Fig.~\ref{fig:size_bc_cluster}. For $v_d<3\,$m/s (approximately), the size 
of the blocking cluster is seemingly of the same order as that of 
the spatial cluster. For $v_d>3\,$m/s, instead, the size of the blocking 
cluster becomes negligible with respect to that of the spatial cluster. \\

We will next turn to study in more detail the size of the spatial cluster. \\

\subsubsection{\label{sec:size_granular}Size of spatial clusters}

\begin{figure*}[!ht]
\subfloat[small and medium spatial 
clusters\label{fig:plot_size_bineado_chicos}]{
\includegraphics[scale=0.715]{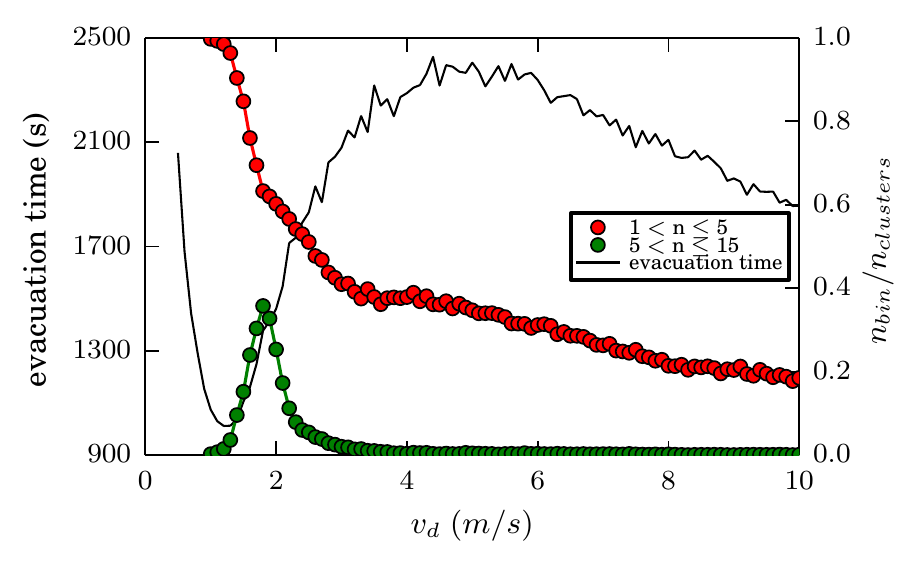}
}
\hspace{-3mm}
\subfloat[big spatial clusters\label{fig:plot_size_bineado_grandes}]{
\includegraphics[scale=0.715]{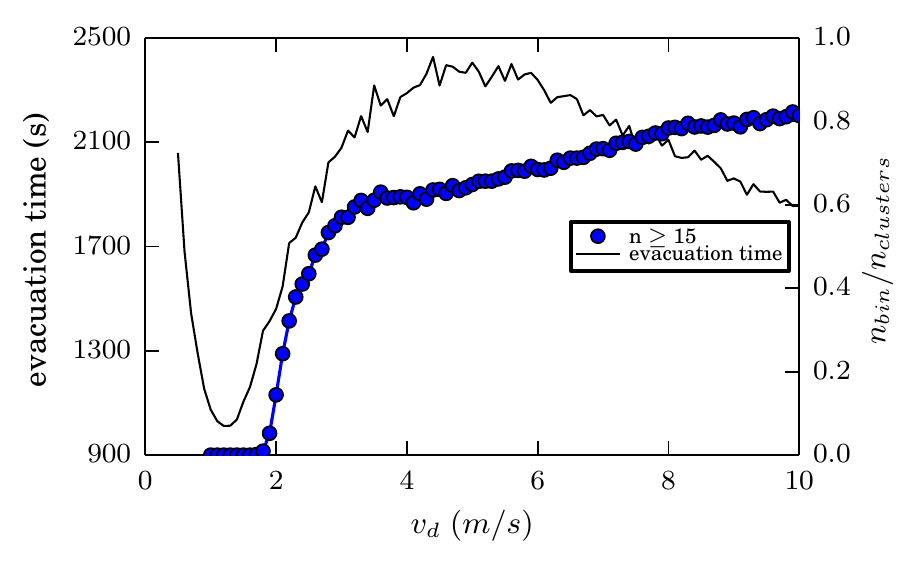}
}
\caption{\label{fig:size_bineado} (a,b) The black line corresponds to the 
evacuation time for the first 7000 evacuees, while the red, 
green and blue circles correspond to the normalized number of spatial clusters 
($n_{bin}$) for three different size categories (see legend for details), as a 
function of the desired velocity $v_d$. The normalization was done with respect 
to the total number of spatial clusters ($n_{clusters}$). The size of each 
spatial cluster was obtained every $0.05\,$s. Data was acquired from a single 
simulation process (re-entering mechanism was allowed).} 
\end{figure*}

Recall that as the desired velocity increases, more people gets into contact 
with each other, and therefore, the size of the spatial cluster increases for 
increasing anxiety levels of the individuals. We classified the spatial clusters 
into three categories (according to their size $n$) as follows: small 
($1<n\leq5$), medium ($5<n<15$) and big ($n\geq15$). We remark that 
the medium category corresponds to the commonly observed size of the blocking 
clusters. Fig.~\ref{fig:size_bineado} shows the (normalized) number of spatial 
clusters for each category (see caption for details). We will analyze each 
category separately. \\

\begin{itemize}
 \item Small spatial clusters ($1<n\leq5$): The occurrence of small spatial 
clusters decreases monotonically as the desired velocity increases. The slope, 
however, appears sharp for $v_d<2\,$m/s and decreases for $v_d>3\,$m/s 
(approximately). The former behavior is complementary to an increment of the 
medium spatial clusters. Also, for $v_d>3\,$m/s the small and big spatial 
clusters complement each other, while no medium clusters are present. An 
inspection of the animations shows that the big spatial cluster always 
surrounds the exit where the pressure maximizes. The small spatial clusters 
appear at the back of the ``bulk''. The latter occur as a result of the 
perturbations when re-injecting pedestrians, and thus, should be considered as 
an artifact from the periodic boundary conditions. \\

\item Medium spatial clusters ($5<n<15$): They occur at the very beginning of 
the \textit{faster is slower} effect and holds for a narrow range of desired 
velocities (say, between $1<v_d<3\,$(m/s)). These are responsible for the 
increase in the evacuation time (as already known), but even before the 
evacuation time arrives to a maximum, the medium size clusters vanishes. \\

\item Big spatial clusters ($n\geq15$): They become significant above 
$v_d=2\,$m/s, approximately. At this threshold, the pressure is high enough to 
force the pedestrians contact each other. A single big cluster would be expected 
to very high desired velocities. \\
\end{itemize}

\subsection{\label{sec:relation_gran_delays}The relation between the size 
of the spatial clusters and the frictional clogging delays}

Recall from Section~\ref{sec:delays_bineado} that we classified the clogging 
delays into three categories: short ($\Delta t \leq1\,$s), intermediate 
(1$\,$s~$<\Delta t \leq3\,$s) and long ($\Delta t >3\,$s). We further 
classified in Section~\ref{sec:size_granular} the size of the spatial clusters 
into other three categories: small ($1<n\leq5$), medium ($5<n<15$) and big 
($n\geq15$). \\

\begin{figure*}[!ht]
\subfloat[$v_d=1.2\,$m/s\label{fig:plot_relation_vd_1_2}]{
\includegraphics[scale=0.67]{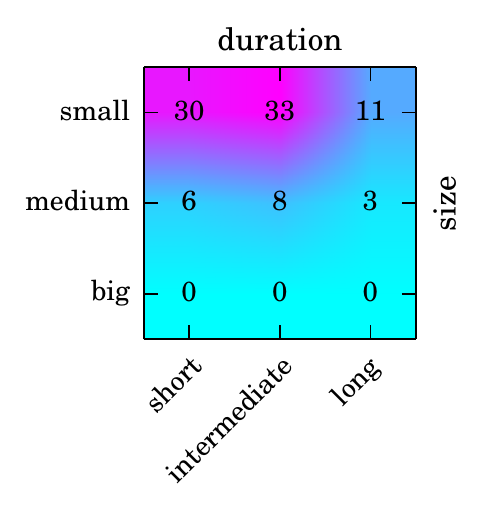}
}
\hspace{-9mm}
\subfloat[$v_d=1.4\,$m/s\label{fig:plot_relation_vd_1_4}]{
\includegraphics[scale=0.67]{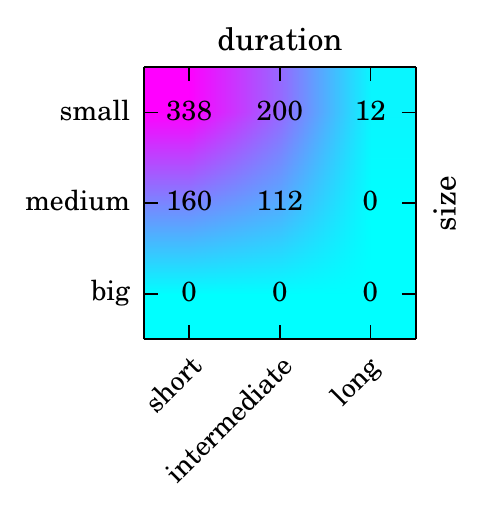}
}
\hspace{-9mm}
\subfloat[$v_d=1.6\,$m/s\label{fig:plot_relation_vd_1_6}]{
\includegraphics[scale=0.67]{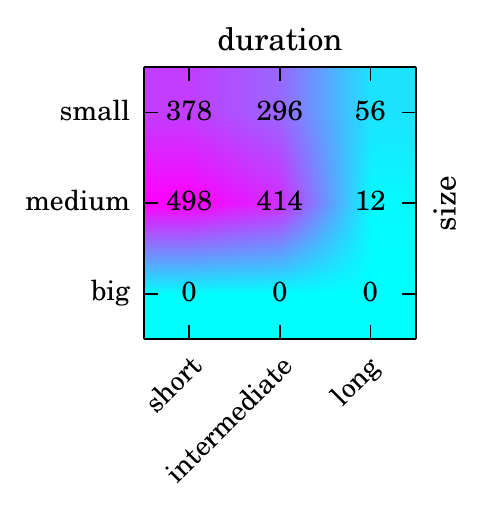}
}
\hspace{-9mm}
\subfloat[$v_d=1.8\,$m/s\label{fig:plot_relation_vd_1_8}]{
\includegraphics[scale=0.67]{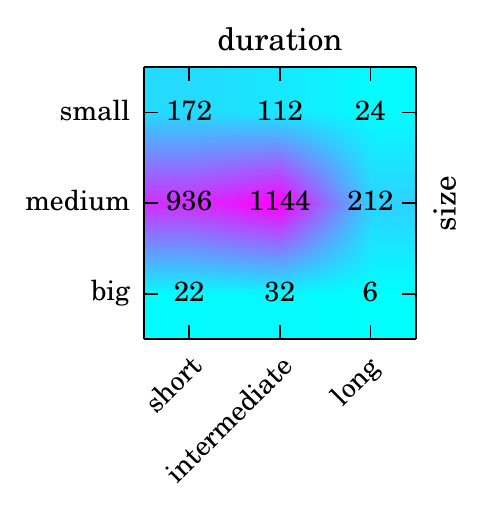}
}
\hspace{-9mm}
\subfloat[$v_d=2.0\,$m/s\label{fig:plot_relation_vd_2}]{
\includegraphics[scale=0.67]{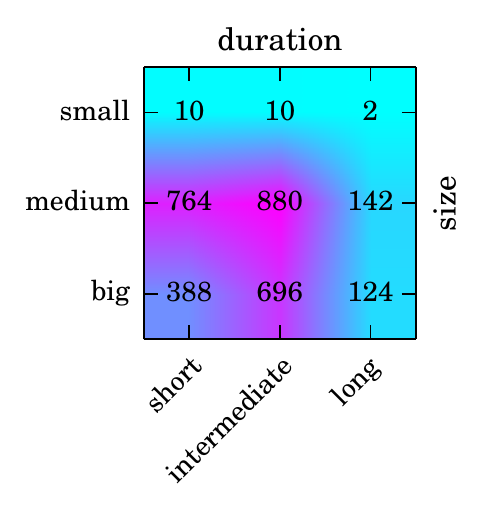}
}
\hspace{-3.5mm}
\subfloat[$v_d=2.2\,$m/s\label{fig:plot_relation_vd_2_2}]{
\includegraphics[scale=0.67]{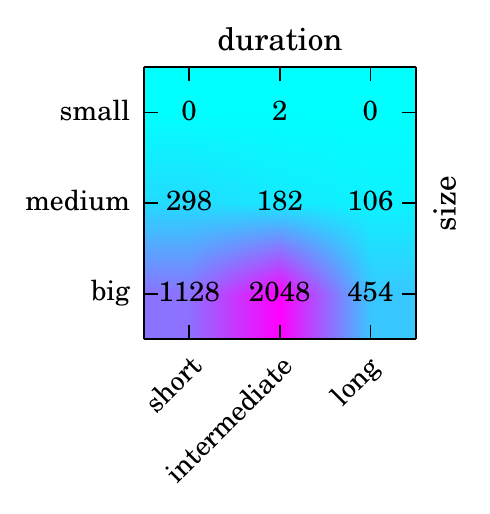}
}
\hspace{-3.5mm}
\subfloat[$v_d=5.0\,$m/s\label{fig:plot_relation_vd_5}]{
\includegraphics[scale=0.67]{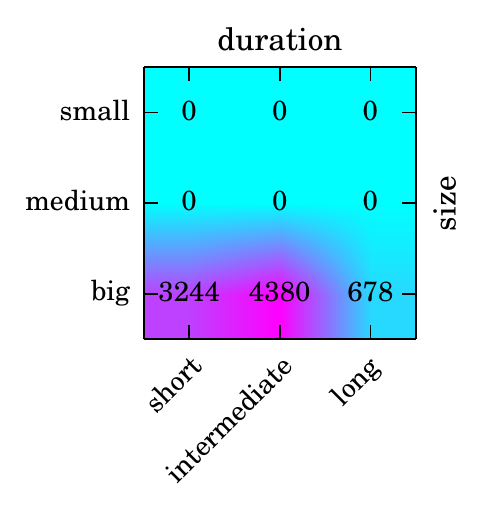}
}
\hspace{-3.5mm}
\subfloat[$v_d=10.0\,$m/s\label{fig:plot_relation_vd_10}]{
\includegraphics[scale=0.67]{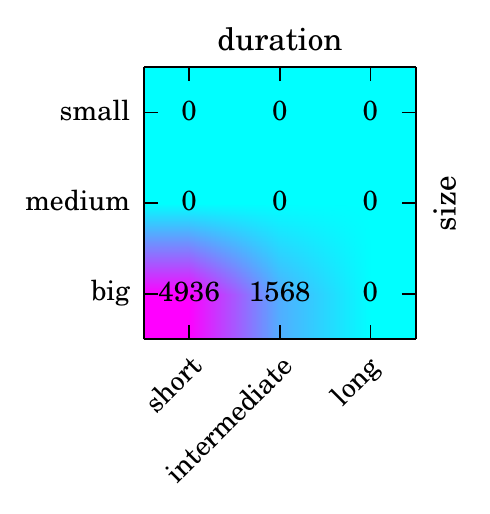}
}
\caption{\label{fig:size_duration} Tile plot between the duration of the 
\textit{frictional clogging delay} and the size of the spatial cluster that 
produces it. In each tile the x-axis represents each of the three clogging 
delay categories and the y-axis represents each of the three size of the 
spatial cluster categories. The number inside each cell corresponds to the 
amount of \textit{frictional clogging delays} that was generated by a specific 
spatial cluster (see text for details).} 
\end{figure*}

We now focus on the relationship between the size of the spatial 
clusters and the corresponding clogging delays. This relationship, however, may 
not be a one-to-one connection since the size of the spatial cluster can change 
between the beginning of a delay and the rupture of a blocking cluster 
(\textit{i.e.} a subset of the spatial cluster). Thus, many spatial clusters 
can be reported during a \textit{frictional clogging delay}. The number of 
spatial clusters (for each category) taking place between the beginning of a 
\textit{frictional clogging delay} and the rupture of the corresponding 
blocking 
cluster is shown in Fig.~\ref{fig:size_duration}. \\

As a first insight, it becomes clear that some kind of correlation between the 
size of the spatial cluster and the associated \textit{frictional clogging 
delay} exists. This correlation is different for different values of the 
desired velocity (see Fig.~\ref{fig:size_duration}). This is in agreement with 
Figs.~\ref{fig:bineado_delays} and \ref{fig:size_bineado}, since both 
magnitudes depend on the desired velocity. Furthermore, we can recognize 
two very different situations by reading the number of reported events inside 
each cell (see caption for details). On the upper row of 
Fig.~\ref{fig:size_duration} (plots (a) to (d)) only small and medium 
spatial clusters are relevant. But, on the lower row (plots (e) to (h)), the 
medium or big spatial clusters are the relevant ones. A closer examination, 
though, shows that a single size is the only relevant one in all the plots 
except for plots (c) and (e) ($v_d=1.6$ and 2$\,$m/s, respectively). \\

The sequence shown in Fig.~\ref{fig:size_duration} for increasing values of 
$v_d$ may be summarized as follows: 

\begin{enumerate}
 \item The patterns (in violet) at $v_d=1.2$ and $1.4\,$m/s correspond to a 
``two box'' flat one. The same patter occurs at $v_d=1.8$ and 2.2 to $10\,$m/s, 
but for different cluster sizes. \\
 \item The ``four box'' square patterns at $v_d=1.6$ and $2\,$m/s are 
transition patterns between two flat ones. \\
\item The small size category is relevant for $v_d<1.6\,$m/s. The medium size 
category becomes relevant for $1.6<v_d<2\,$(m/s). The big spatial clusters 
are significant for $v_d>2\,$m/s.
\end{enumerate}

Interestingly, Fig.~\ref{fig:size_duration} also shows that either the three 
categories of frictional clogging delays and the spatial cluster sizes 
are present for $v_d=1.8$ and $2\,$m/s (all the cells are not empty). This 
means that the categories are somehow ``mixed'' at these desired velocities. We 
can find small, medium and big spatial clusters of all delays long. \\

Our major conclusions from this Section are as follows. First, two kinds of 
spatial clusters are relevant within the \textit{faster is slower} and 
\textit{faster is faster} regimes. From the point of view of the spatial 
clusters, small and medium spatial clusters ($n<15$) dominate the scene for 
desired velocities below $3\,$m/s. These (almost always) match the blocking 
cluster definition. But, for $v_d>3\,$m/s, the giant spatial 
cluster ($n\geq15$) becomes the relevant structure during the evacuation 
process. We may associate the overall delays (for $v_d>3\,$m/s) to the existence 
of this structure rather than its inner most perimeter only (say, the blocking 
cluster). \\

As a second conclusion, we noticed that (approximately) short and intermediate 
frictional clogging delays can be associated with small spatial clusters for 
low desired velocities (say, less or equal to $1.4\,$m/s). Besides, these 
clogging delays can be associated with big spatial clusters for desired 
velocities greater than $3\,$m/s. The situation in between, say 
$1.4<v_d<3$\,(m/s), is somewhat mixed: two ``transitions'' occur at $v_d=1.6$ 
and 2$\,$m/s (approximately). The former corresponds to a 
small$\,\rightarrow\,$medium stage, while the latter corresponds to a 
medium$\,\rightarrow\,$big stage. All these regimes attain small/intermediate 
clogging delays. \\ 

\section{\label{sec:conclusions}Conclusions}

Our research focused on the microscopic analysis of the evacuation process of 
self-driven particles confined in a square room with a single exit door. We 
simulated 225 individuals escaping through a door with a width of two times a 
pedestrian width. The simulations were done in the context of the Social Force 
Model. We were mainly interested in properly understanding the 
\textit{faster-is-slower} and \textit{faster-is-faster} phenomena, 
\textit{i.e.} the relation between the flow of pedestrians and the formation 
of structures (clusters) which might impede the motion of the walkers.

We have found that as the desire velocity increases, the system evolves 
from a condition in which the flow decreases with increasing $v_d$ to another 
in which the flow increases with increasing $v_d$. The cause of such a change 
can be traced to the characteristics of the above mentioned blocking structures. 
We have shown that the \textit{faster-is-slower} and the 
\textit{faster-is-faster} are quite different phenomena from the standpoints of 
the ``frictional'' clogging delays and the underlying clusterization structure. 
The \textit{faster-is-slower} occurs whenever the blocking clusters dominate 
the scene, \textit{i.e.} there is a structure which anchors to the walls of the 
exit and is able to momentarily obstruct the exit, accomplishing moderate to 
long lasting clogging delays (say, above 1$\,$s). However, as the $v_d$ 
increases, the pressure increases as well and then, the blocking cluster 
becomes a giant cluster, involving (almost) all the crowd. The giant cluster is 
not anchored to the walls but resembles a very viscous fluid. Keep in mind that 
in the Social Force Model the agents are not represented by hard spheres but by 
soft spheres and  as such the resistance opposed by the agents can be overcome 
by a force big enough. The harder the pedestrians push, the weaker the 
slowing-down, achieving the \textit{faster-is-faster} phenomenon. This 
phenomenon has not been reported on other granular systems, to our knowledge. \\

For the set of parameters of the Social Force Model used in this work the 
analysis of the correlation between the clogging delays and the spatial 
structures showed that for $v_d>5\,$m/s, only short ($\Delta t \leq1\,$s) and 
intermediate (1$\,$s~$<\Delta t \leq3\,$s) clogging delays are produced during 
the giant cluster scenario. This means that the pushing efforts overcome the 
friction between pedestrians belonging to the ``blocking'' cluster, and thus,  
this kind of structures are not able to hold for a very long time (say, longer 
than 3$\,$s). The blocking dynamic is replaced then by the collective dynamic of 
the giant cluster. Curiously, we noticed that short delays become increasingly 
relevant during the \textit{faster-is-faster} scenario, since the burst of 
leaving pedestrians tends to be increasingly regular. \\

A by-product of our investigation is that we can distinguish between two 
categories of clogging delays, according to their production mechanism. The 
social clogging delays corresponds to those that are generated as a consequence 
of the social force among individuals. Instead, the frictional clogging delays 
are a consequence of the granular force among individual. However, social and 
frictional clogging delays are complementary delays whenever a blocking cluster 
breaks down, releasing a burst of individuals. We should call the attention on 
this point when analyzing the overall delay of an evacuation.

\section*{Acknowledgments}
This work was supported by the National Scientific and Technical 
Research Council (spanish: Consejo Nacional de Investigaciones Cient\'\i ficas 
y T\'ecnicas - CONICET, Argentina) and grant Programaci\'on Cient\'\i fica 2018 
(UBACYT) Number 20020170100628BA. G. Frank thanks Universidad Tecnol\'ogica 
Nacional (UTN) for partial support through Grant PID Number SIUTNBA0006595.

\appendix

\section{\label{sec:K-S}Statistical analysis of the discharge curves}

This analysis was carried out by means of the Kolmogorov-Smirnov test. This 
non-parametric statistical test is a goodness of fit test, which allows 
comparing the distribution of an empirical sample with another, or, to what is 
expected to be obtained theoretically (null hypothesis) \cite{frodesen}. \\

The following statistic is considered

\begin{equation}
D_n = \mathrm{max} \left | S_n(x) - F_o(x) \right | \label{eq:k-s}
\end{equation}

\noindent where S$_n(x)$ and F$_o(x)$ represent the cumulative distributions of 
the sampled data and the theoretical one, respectively. As can be seen, this 
test quantifies the maximum (absolute) difference between both distributions.\\

In our case, the discharge curves are compared against a uniform flow of 
individuals. The latter corresponds to delays of the same duration. This is 
useful for analyzing the relative significance of the delays during the 
evacuation process.\\

\begin{figure}[!ht]
\centering
\includegraphics[width=0.7\columnwidth]{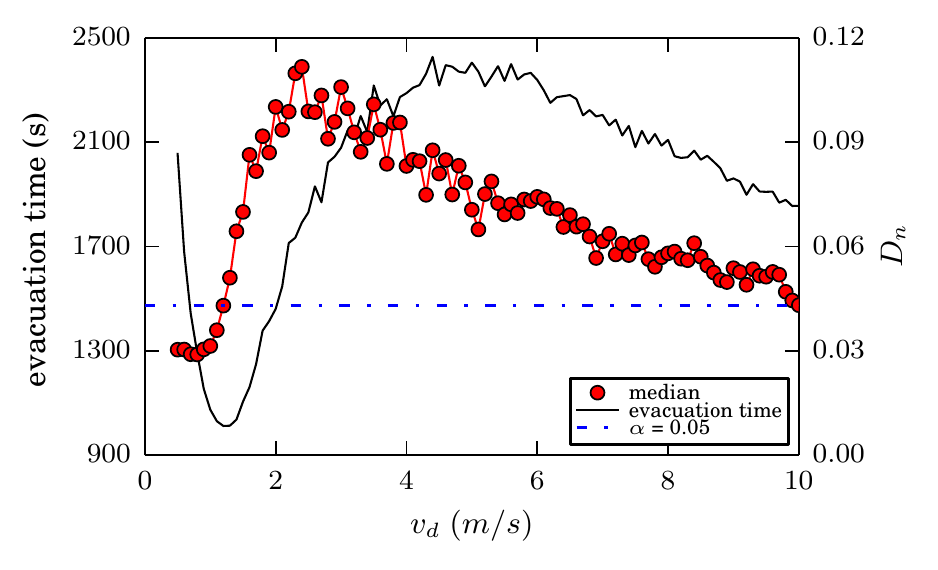}
\caption{\label{fig:kolmogorov} The black line corresponds to the 
evacuation time for the first 7000 evacuees as a function of the desired 
velocity $v_d$. Red circles correspond to the Kolmogorov-Smirnov test 
(see text for details). The blue horizontal dashed line corresponds to 
a significance level of $5\%$. 40 discharge curves were analyzed for each 
desired velocity Each discharge curve corresponds to the evacuation process of 
180 individuals. These were uncorrelated sub-intervals from a single 
simulation process (re-entering mechanism was allowed).}
\end{figure}

To carry out the Kolmogorov-Smirnov test, 40 discharge curves were analyzed, 
each of which correspond to the evacuation process of 180 individuals. The 40 
selected processes were chosen separately in time, in order to be 
uncorrelated. The results are presented in Fig.~\ref{fig:kolmogorov}.\\

As can be seen, the behavior of the median coincides qualitatively with that 
corresponding to the evacuation time. However, both curves are horizontally 
shifted from each other. The maximum of the curve corresponding 
to the median is approximately at $v_d=2\,$m/s, unlike what happens with the 
evacuation time ($v_d=5\,$m/s). It should be noted that the former occurs in 
the moment of the inversion of the concavity of the curve associated with the 
evacuation time.\\

We can identify two regimes from Fig.~\ref{fig:kolmogorov}, depending on the 
desired velocity. For $v_d<2\,$m/s, the discrepancy between the simulated and 
uniform flow increases as the desire velocity increases. This means that the 
flow becomes more irregular as the anxiety of the individuals increases. As will 
be seen in Section~\ref{sec:clusters}, this is due to the influence of blocking 
clusters during the evacuation process.\\

On the other hand, for $v_d>2\,$m/s, the discrepancy $D_n$ as the 
desire to escape from the room increases. Thus, the flow becomes more regular. 
It is worth mentioning that this behavior starts during the \textit{faster is 
slower} effect (and continues during the \textit{faster is faster} effect). In 
addition, it is possible to notice a linear behavior of the median in this range 
of desired velocity.\\

Finally, in order to quantify the discrepancy between both distributions, a 
significance level of $5\%$ is indicated in Fig.~\ref{fig:kolmogorov} (dashed 
line). As can be seen, the null hypothesis is rejected for approximately the 
entire explored range of desired velocity. On the other hand, it should be 
noted that this is not possible before the minimum evacuation time 
(\textit{i.e.} $v_d<1.1\,$m/s). At this interval, the individuals are actually 
not in contact. Therefore, the flow of evacuees occurs uninterruptedly due to 
the absence of blocking clusters. Thus, we can only concluded that the flow it 
is not uniform.\\

\section{\label{sec:prob_bc}Blocking cluster probability}

\begin{figure}[!ht]
\centering
\includegraphics[width=0.7\columnwidth]{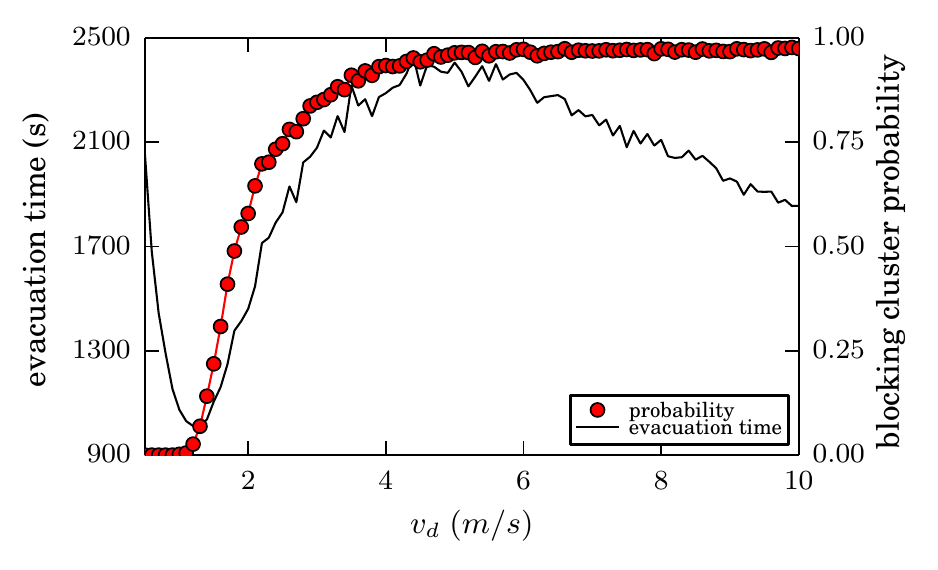}
\caption{\label{fig:probability_bc} The black line and red circles correspond 
to the evacuation time for the first 7000 evacuees and the probability of 
a blocking cluster as a function of the desired velocity $v_d$, respectively. 
These were computed from a single simulation process (re-entering mechanism 
was allowed).}
\end{figure}

We analyze here the probability of occurrence of a blocking cluster (see 
Fig.~\ref{fig:probability_bc}). That is, the percentage of time that the system 
is in the presence of a blocking cluster. Recall that in 
Fig.~\ref{fig:variando_tc} it was observed that, for any given delay value, as 
the desired velocity increases, so does the probability that this delay 
was generated by a blocking cluster.\\

As can be seen, the higher the desired velocity, the greater the probability of 
attaining a blocking cluster. Moreover, it has an asymptotic behavior 
above $v_d=5\,$m/s, approximately. This means that above this desired 
velocity, the presence of blocking clusters is permanent over time. However, we 
can note that despite the still presence of a blocking cluster in front of 
the door, evacuation time decreases. The explanation for this phenomenon are 
explained in Section~\ref{sec:clusters}.\\

The above results are in agreement with those reported in 
Refs.~\cite{dorso_2007,dorso_2005}, in spite of the different boundary 
conditions, ranges of desired velocity and/or the value of the elastic constant 
$k_n$ \cite{Sticco_2020}. But, unlike the results reported in the literature 
where this behavior occurs in the presence of the \textit{faster is slower} 
effect, in this case we observe that it is also satisfied during the 
\textit{faster is faster} effect.\\

Besides, we were able to observe through the process animations that at the 
moment in which a blocking cluster is fractured, it is replaced by another 
immediately for $v_d>5\,$m/s (\textit{i.e.} probability of blocking cluster 
equal to one). Also, the videos show that the number and composition of 
individuals that make up the blocking cluster changes dynamically over time. 
However, despite this, the presence of a blocking cluster in front of the door 
is permanent for $v_d>5\,$m/s.\\

\section*{References}

\bibliographystyle{elsarticle-num}
\bibliography{paper}

\end{document}